\documentclass[compsoc, conference, letterpaper, 10pt, times]{IEEEtran}

\usepackage{amsmath,amssymb,array,color,colortbl,graphicx,multirow}
\usepackage{microtype}
\usepackage[pdfpagelabels=false,bookmarks=false]{hyperref}
\usepackage{url}

\usepackage[anythingbreaks]{breakurl}
\usepackage{comment}

\usepackage{algorithm2e}
\usepackage{comment}
\usepackage{listings,color}
\usepackage{verbatim}
\usepackage{multirow}
\usepackage{booktabs}
\usepackage{balance}

\usepackage{listings}

\usepackage{framed,color}

\usepackage{graphicx}
\usepackage{caption}
\usepackage{subcaption}

\usepackage[para]{footmisc}

\newcommand{\oobf}{\emph{Out-of-Band Forwarding}}

\definecolor{mygreen}{rgb}{0,0.5,0}

\newcommand{\stefan}[1]{\textit{\textcolor{red}{[stefan]: #1}}} 
\newcommand{\kash}[1]{\textit{\textcolor{blue}{[kash]: #1}}} 
\newcommand{\Hilightgreen}{\makebox[0pt][l]{\color{green}\rule[-4pt]{.10\linewidth}{14pt}}}
\newcommand{\Hilightred}{\makebox[0pt][l]{\color{red}\rule[-4pt]{.10\linewidth}{14pt}}}

\begin{document}

\title{Outsmarting Network Security with SDN Teleportation}

\author{}

\author{
 Kashyap Thimmaraju$^{1}$ \quad Liron Schiff$^2$ \quad \quad Stefan Schmid$^{3}$\\
{\small $^1$ TU Berlin, Germany \quad 
\quad $^2$ GuardiCore Labs, Israel 
\quad~$^3$ Aalborg University, Denmark}
}
\author{\IEEEauthorblockN{Kashyap Thimmaraju}
\IEEEauthorblockA{Security in Telecommunications\\
TU Berlin\\
Berlin, Germany\\
Email: kash@sect.tu-berlin.de}
\and
\IEEEauthorblockN{Liron Schiff}
\IEEEauthorblockA{GuardiCore Labs\\
Tel Aviv, Israel\\
Email: liron.schiff@guardicore.com}
\and
\IEEEauthorblockN{Stefan Schmid}
\IEEEauthorblockA{Dept. of Computer Science\\
Aalborg University\\
Aalborg, Denmark\\
Email: schmiste@cs.aau.dk}}


\maketitle


\begin{abstract}
Software-defined networking is 
considered a promising new
paradigm, enabling more reliable 
and formally verifiable communication
networks. 
However, this paper shows that 
the separation
of the control plane from the data plane,
which lies at the heart of Software-Defined Networks (SDNs),
introduces a new vulnerability which we call 
\emph{teleportation}.
An attacker (e.g., a malicious switch 
in the data plane or a host connected
to the network)
can use teleportation
to 
transmit 
information
via the control plane and
bypass critical network functions in the data plane 
(e.g., a firewall),
and to  
violate security policies as well as 
logical and even physical
separations.  
This paper characterizes the design space
for 
teleportation attacks theoretically, and then
identifies
four different teleportation techniques.
We demonstrate and discuss how these techniques 
can be exploited for different attacks
(e.g., exfiltrating confidential data
at high rates), and also 
initiate the discussion of 
possible countermeasures.
Generally, and given today's trend toward
more intent-based networking, we believe that our findings are 
relevant beyond the use cases considered in this paper.
\end{abstract}

\section{Introduction}\label{sec:intro}

Computer networks such as datacenter
networks or the Internet
have become
a critical infrastructure~\cite{eu}. Not only a large fraction of
the economic activity critically depends on the availability
of such networks, but also governments increasingly rely on
existing and shared infrastructures, mainly for their cost benefits~\cite{cheap}.

This dependency on public and shared infrastructures 
raises concerns. While the Internet has certainly been
a
huge success, and over the last years, there has been much innovation on the higher network
layers (e.g., application layer) and the lower network layers (e.g., data-link and physical layer),
the Internet core suffers from ossification~\cite{ossi}.
In particular, it is questionable whether today's network technology
is sufficient to ensure essential security, resilience and dependability properties. 
For instance, today's Internet does not provide
any means of path control, and we are still struggling to render
routing protocols more secure~\cite{psbgp}.

Software-Defined Networking is a novel
networking paradigm which 
promises to enable these necessary
innovations, also in terms of security,
through its decoupling and consolidation of the control plane, its 
formally verifiable  
policies~\cite{header-space,veriflow,sharon,consistent-updates},
as well as by introducing new functionality~\cite{netlord,fortnox,fresco,avantguard}.
 
\begin{figure}[t]
    \begin{center}
        \includegraphics[width=0.99\columnwidth]{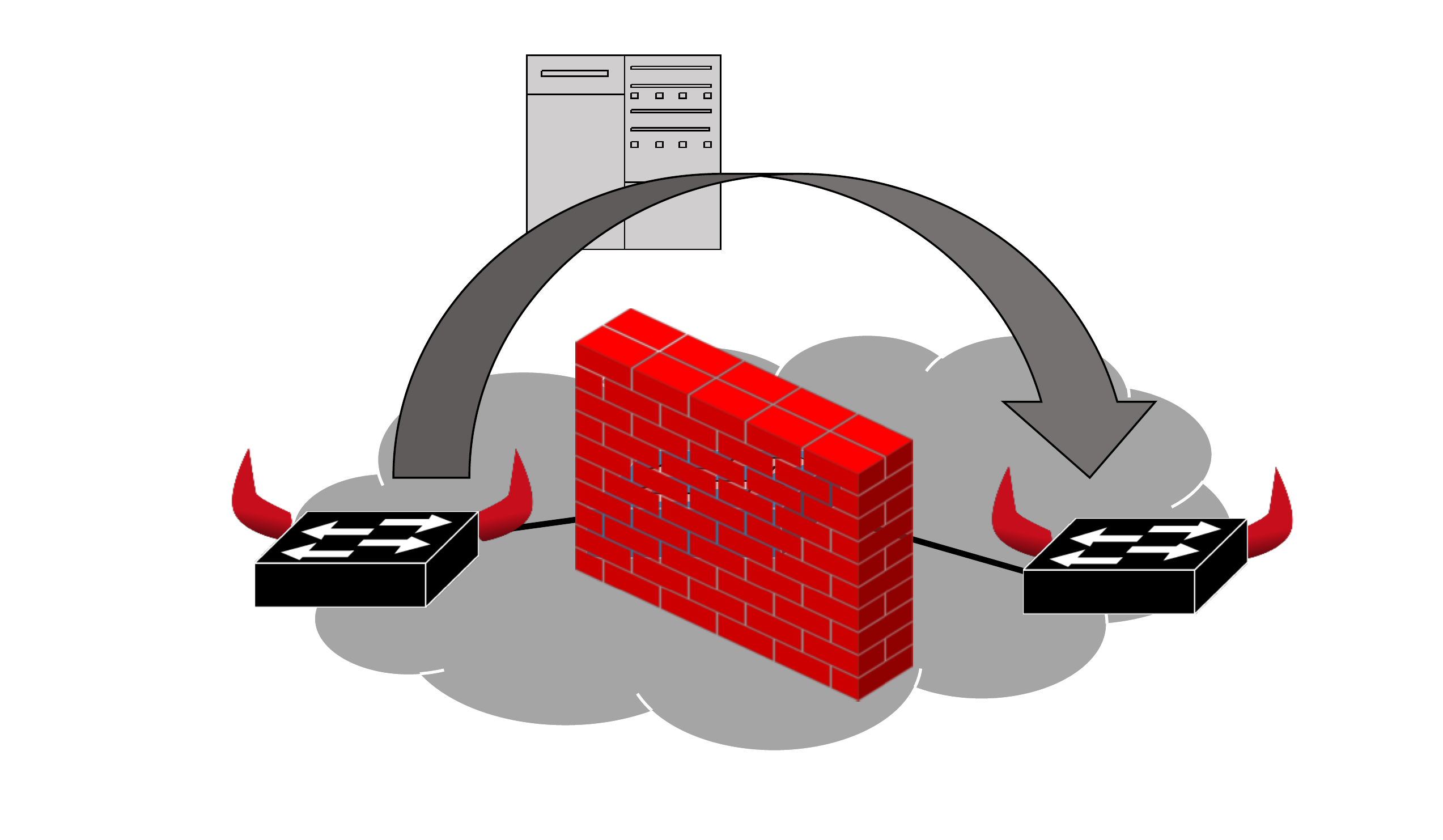}\\
    \end{center}
    \caption{Illustration of teleportation:
     Malicious switches
(with \emph{red horns})
exploit the control platform for hidden communication,
possibly bypassing data plane security mechanisms
such as a firewall.}
    \label{fig:covert}
\end{figure}

However, Software-Defined Networks (SDNs) also introduce
new security challenges.
In particular, we in this paper study threats
introduced by an \emph{unreliable south-bound interface},
i.e., we consider a threat model in which 
switches or routers
do not behave as expected, but rather are malicious~\cite{thimmaraju2016reigns}, and e.g.,
contain hardware backdoors~\cite{snowdencisco}. 
While many existing network security and monitoring tools rely on 
the trustworthiness of switches and routers,
this assumption has become questionable:
Attackers have repeatedly demonstrated their ability to compromise
switches and routers~\cite{1001dsl,synful, lindner2009cisco},
thousands of compromised access and core routers are 
being traded underground~\cite{lee2006secure}, 
networking vendors have left backdoors open~\cite{huawei, netisbackdoor},
national security agencies
can bug
network equipment~\cite{snowdencisco},
hacker tools to scan and eventually exploit routers with weak passwords,
default settings
 are openly available on the Web, etc. 
However, the impact of malicious hardware is not well understood
and underexplored today.

In particular, this paper shows how an outsourced and consolidated 
control plane---as it lies at the heart of the SDN paradigm, introduces
an opportunity for
\emph{teleportation}: Malicious SDN switches
may transmit information  
via the logically centralized software control plane,
\emph{completely} bypassing 
data plane elements (such as other switches, middleboxes, etc.).
By violating logical and even physical separations, teleportation
can constitute a serious security threat.
For example, teleportation could be used by 
one malicious switch to discover (and communicate information to) other 
malicious switches, bypassing security checks in the data plane. 
As we will show in this paper, teleportation can also be 
exploited by malicious hosts, triggering (benign) switches 
to teleport information for them.

We argue that teleportation can 
be seen as a flexible communication channel which
constitutes a threat in various situations, for example
(see Figure~\ref{fig:covert}):
\begin{enumerate}
\item \textbf{Bypassing critical network components:} 
By implicitly communicating information 
via the control plane, it is possible to circumvent
critical network components, such as switches, middleboxes
or policy enforcers located in the data plane. For example,
teleportation can in principle be used to bypass
middleboxes performing security checks 
(e.g., network intrusion detection
systems), 
middleboxes in charge of billing (e.g., a Radius server),
or QoS enforcers (e.g., a leaky bucket policer).

\item \textbf{Rendezvous and attack coordination:}
While already a single malicious switch, for example located \emph{inside} an
enterprise network, may cause significant harm and violate
basic security policies, the situation becomes worse
if multiple malicious switches cooperate~\cite{inception}.
Malicious or Trojan switches (e.g., switches containing
a hardware/software backdoor) may use teleportation as
a rendezvous protocol, to discover each other, and 
subsequently coordinate
an attack.

\item \textbf{Exfiltration:}
Teleportation can also be used to exfiltrate
sensitive information between 
networks that have no data plane connectivity.

\item \textbf{Eavesdropping and data tampering:} 
Particularly serious threats are introduced if malicious hosts and switches collude.
For instance, in a scenario
with collusion,
teleportation can be used for eavesdropping. We show that  
a malicious switch and host can
carry out a man-in-the-middle attack
that serves benign hosts with malicious
web pages.
\end{enumerate}

Teleportation can be 
difficult to detect:
The teleported information follows the normal traffic pattern of 
control
communication, not between switches directly but \emph{indirectly}
between any switch and the controller. 
Moreover,  
the teleportation channel
is \emph{inside}
the (typically encrypted) OpenFlow channel.
Accordingly, it cannot easily be detected with 
modern Network Intrusion Detection Systems (NIDSs), 
even if they operate in the control plane.

\subsection{Our Contributions}

This paper 
makes the following contributions:
\begin{enumerate}
\item We identify a 
new vulnerability, namely \emph{teleportation}, which targets the very core of the software-defined
networking paradigm, namely the separation of the control
and the data plane. 
In particular,
we consider the threats introduced
by a malicious data plane.
Indeed,
recent incidents related to the 
trustworthiness of routers and switches, indicate that our
threat model is a relevant one.

\item We model and characterize
possible teleportation
channels theoretically.

\item We recognize and demonstrate
four teleportation techniques 
in software-defined networks utilizing
state-of-the-art OpenFlow controllers,
in particular
ONOS~\cite{onos} among others.

\item We present and demonstrate multiple
simple and sophisticated attacks. 
In particular, we 
show that teleportation can also be exploited by malicious hosts in scenarios
where all switches are benign. 

\item We evaluate the performance and quality of
the teleportation channel in terms of
throughput, jitter and packet loss respectively,
and also evaluate the resource footprint in terms of
CPU and memory consumption at the controller.

\item We initiate the discussion of possible countermeasures. 
In particular, we propose to combine intrusion
detection with waypoint-enforcement, 
forcing 
\emph{Packet-out} messages (from controller to switches)
to pass through the waypoint if mandated
by a security policy.


\end{enumerate}

We have already notified the open source community
about some of the issues reported in this paper,
and first actions have been taken
(see CVE-2015-7516~\cite{onosDos}).

More generally, in the light
of today's trend toward more intent-based networking, we believe that our work touches
a topic whose relevance may increase in the near future and go beyond
the use cases considered in this paper. 

\subsection{Paper Organization}

The remainder of this paper is organized as follows.
Section~\ref{sec:background} introduces the necessary background
on OpenFlow and SDN.
Section~\ref{sec:model} introduces our threat model, and
Section~\ref{sec:covert} characterizes possible
teleportation channels. Section~\ref{sec:secblocks} 
describes teleportation techniques;
based on these techniques, 
we demonstrate and discuss different attacks in Section~\ref{sec:seceval}.
Section~\ref{sec:perf} describes our performance evaluation
of the out-of-band forwarding channel.
Section~\ref{sec:possible} initiates the discussion of countermeasures.
After reviewing related work in 
Section~\ref{sec:relwork},
we conclude our work in Section~\ref{sec:conclusion}.

\section{Preliminaries}\label{sec:background}

This paper considers Software-Defined Networks (SDNs)
which outsource and consolidate
the control over the network switches to a logically centralized
software controller. The separation of the control and data plane
has the potential to simplify the network management, 
as many networking
tasks are inherently non-local. Moreover, SDN and especially
its de facto standard protocol, OpenFlow, 
introduce interesting new flexibilities, e.g., in terms of 
traffic steering: Routes may not necessarily be destination based,
and can depend on layer-2, layer-3 and even layer-4 properties
of the packets. 

At the heart of an 
SDN lies a control software, running on a
set of servers. These controllers receive information and
statistics from switches, and depending on this information
as well as the policies they seek to implement, issue
instructions to the switches.

OpenFlow follows
a match-action paradigm: The controllers install rules
on the switches which consist of a match and an action
part; the packets (i.e., flows) matching a rule are subject
to the corresponding action. 
That is, each switch stores
a set of (flow) tables which are managed by the controllers, and
each table consists of a set of flow entries which specify
expressions that need to be matched against the packet
headers, as well as actions that are applied to the packet
when a given expression is satisfied. Possible actions
include dropping the packet, sending it to a given egress
port, or modifying its header fields, e.g., adding a tag. 
The match-action
paradigm is attractive as it simplifies formal reasoning 
and enables policy verification.

By default, if a packet arrives at a switch
and does not match an existing 
rule, the packet (usually without payload if the switch supports packet buffering)
is forwarded to the controller.
This event is called a \emph{Packet-in}.
Upon a \emph{Packet-in} event, the controller can decide how to react
to packets of the corresponding type, and add/delete/modify
flow rules accordingly issuing \emph{Flow-mod} 
messages to the switch (and maybe to other switches proactively on this occasion 
as well).
A controller can also decide to send out a packet explicitly
from a switch, issuing a so-called \emph{Packet-out}
command to the switch.

An attractive alternative to the hop-by-hop installation
of new flows, reacting  to a new packet repeatedly along the path (multiple \emph{Packet-in}s), 
is the so-called ``pave-path technique'': Once the controller receives a first 
\emph{Packet-in} event from some switch, it proactively updates the other
switches along the path.
Such an ``intent-based'' controller behavior can 
render the reaction to network events and set up
of host-to-host/network connectivity (according to current policies)
more efficient.

While SDNs are logically centralized, the control plane can be
physically distributed, e.g., for fault-tolerance or performance reasons.
Accordingly, OpenFlow supports 
multiple controllers for a single switch.
The controllers and switch exchange \emph{Role-request}
and \emph{Role-reply} messages respectively to assert
the various roles (\emph{Master}, \emph{Equal} and 
\emph{Slave}).
There may be only one \emph{Master} controller for a given
switch while multiple \emph{Equal} and \emph{Slave} controllers
are permitted.

The OpenFlow standard~\cite{specification2013version} specifies
basic security mechanisms. For example,
the communication between the controller and switch
can be authenticated and encrypted, using
TLS over TCP/UDP.

Finally, we note that although some of our techniques are generally
applicable in networks separating the control plane and the data plane,
while others exploit OpenFlow specific features, when clear from the context,
in this paper we will treat SDN and OpenFlow
as synonyms.

\section{Threat Model}\label{sec:model}

We in this paper consider a threat model where 
OpenFlow switches, hosts, or both, may not behave 
correctly but are malicious. 

We do not place any restrictions on what a 
malicious switch can and cannot do.
For example, a  malicious switch can fabricate and transmit 
any type of OpenFlow message, it can arbitrarily deviate 
from the OpenFlow specification, and it can even  
use multiple identities, 
all at the risk of being detected.
However, the malicious switch cannot choose
where it will be placed in the network.
In order to collude, the malicious switches have been programmed 
to recognize some data and/or timing pattern. 
%
%
Similarly, we do not place any restrictions on what a malicious host
can and cannot do. For example, a malicious host
may masquerade its Media Access Control (MAC) and/or 
Internet Protocol (IP) addresses, use
an incorrect gateway, falsify Address Resolution Protocol (ARP) requests/responses,
and so on.  
The attacker could also be an insider, i.e.,
an authorized user
who intends to subvert his/her current organization.
We also consider the case where malicious hosts and
switches collude.
We assume that an attacker has sufficient
resources and know-how to compromise hosts/switches and therefore 
do not
concern ourselves with
how the host/switch is compromised.
For example, the attacker can exploit a buffer overflow vulnerability
in the switch software to compromise
the switch~\cite{thimmaraju2016reigns}.

The OpenFlow controller and its applications
on the other hand are trusted entities
and are available to the switches:
For example, they are based 
on static and
dynamic 
program analyses.
The OpenFlow channel
is reliable and may be encrypted.

\section{Modeling Teleportation}\label{sec:covert}

With these concepts in mind, we now model and characterize 
a novel threat called \emph{teleportation} which targets the heart of SDNs:
The outsourcing and consolidation of control over multiple data plane elements.
In particular, we argue that we can see 
an OpenFlow controller as a 
``reactor'': It reacts (in a best-effort and timely manner) to events 
generated by the network operator, the OpenFlow switches, 
and timeouts; as a response, 
the controller sends OpenFlow commands to switches. 
Accordingly, we argue that the following 3-stage functionality 
is fundamental in the SDN paradigm. 
\begin{enumerate}
\item \textbf{Switch to controller:} A source switch 
intentionally or unintentionally sends modulated information 
to the controller (e.g., by adding specific events, delaying existing events, etc.).
\item \textbf{Controller to switches:}
The controller reacts to the received events, by sending commands to 
one or multiple other switches.
\item \textbf{Destination processing:}
 A destination switch processes incoming commands 
 from the controller. 
 In case of a malicious switch,
 the switch may search for 
 some message properties, temporal patterns, etc.,  and 
 hence infer
 the information modulated by the source,
 or by simply forwarding the information (to a potentially malicious host). 
\end{enumerate}

Based on this controller model, we can identify two kinds of 
teleportation channels~\footnote{We note that our terminology of teleportation can be viewed as
analogous to covert channels. Explicit teleportation is
analogous to covert storage channels and implicit teleportation is
analogous to covert timing channels.}:
\begin{itemize}
\item \textbf{Explicit teleportation:} The teleported information actually appears in the messages
exchanged. The message may for example contain steganographic contents.
\item \textbf{Implicit teleportation:} The teleportation 
relies on modulating information implicitly. For example, it is based on timing
(e.g., message transmissions are delayed according to some pattern)
or it is based on shared resources, whose
availability is changed over time (e.g., leveraging mutual exclusion). 
\end{itemize}

\section{Teleportation Techniques}\label{sec:secblocks}

Having established a conceptual model of teleportation,
we next present techniques that can realize teleportation in
today's SDNs. 
In particular, we have identified the following three 
fundamental SDN functionalities which 
can be
exploited for teleportation:

\begin{enumerate}
\item \textbf{Flow (re-)configurations:} In an SDN,
a controller needs to react to various data plane events (such as so-called 
\emph{Packet-in}s
in OpenFlow
or link failures), and configure and reconfigure flows and paths accordingly.
Triggering and exploiting such events can be used for teleportation.

\item \textbf{Switch identification:} In an SDN, 
switches are responsible for introducing
and uniquely identifying themselves to the controller. This is required as policies are
often specific to the switch. Unique switch identifiers are also necessary to
correctly construct and enforce policies on the switches and in the controller.
We will show that such switch identification mechanisms can be exploited for
teleportation.

\item \textbf{Out-of-band forwarding:} An SDN controller
 must not only be able to receive events and control packets from switches,
 but also to instruct switches to forward specific messages. 
 This basic functionality in SDNs can be exploited by a malicious
 switch or host to forward entire
 packets via the controller.
\end{enumerate}

In the remainder of this section, we will discuss these teleportation techniques
in more detail in turn.

\subsection{Flow (Re-)Configurations}\label{sec:tele_path_update}

We distinguish between two types of flow
re-configuration events: \emph{path update} and 
\emph{path reset}.

\textbf{Path Update:}
Our first teleportation technique is based on \emph{path updates}.
Path updates are a fundamental controller functionality,
and come in the form of different controller features
such as \emph{Mobility}, \emph{VM Migration}
or simply \emph{MAC Learning}. 
The basic scheme is as follows: A controller
typically maintains some mapping of which hosts (MAC addresses)
are connected to which ports (on the switch). If a host suddenly appears on 
another switch, the controller installs new flows for the host 
on the new switch, and also deletes the corresponding flow rules 
on the old switch. We define this type of installation and deletion
of flows by the controller on switches as \emph{path update}. Specifically,
a path update involves the use of \emph{Packet-in}, \emph{Flow-mod} 
and \emph{Packet-out}
messages. Malicious switches can use path update for
implicit teleportation.

\begin{figure}[t]
    \begin{center}
		\small
		\def\svgwidth{3.187in}
        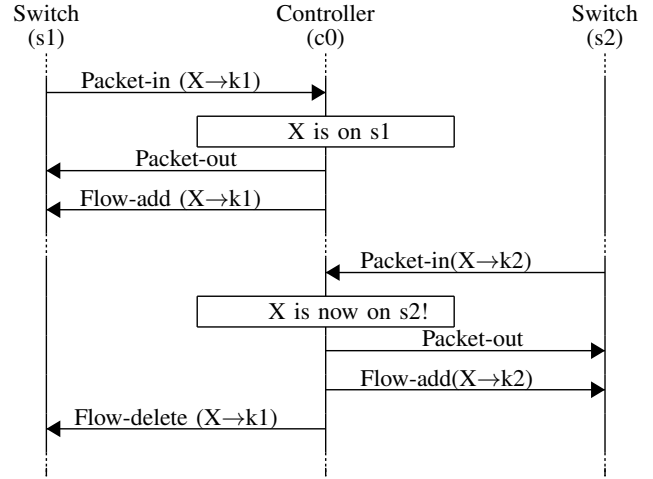
    \end{center}
    \caption{Message sequence pattern for path update teleportation. Switch
   ~$s2$ teleports information to~$s1$ when~$s1$ receives the Flow-delete message from controller~$c0$.}
    \label{fig:hm_onos}
\end{figure}

For the teleportation with path update, 
a switch 
triggers the deletion of rules at other switches. 
Malicious switches can teleport information between themselves
by prompting path updates for the same host using
\emph{Packet-in}s. Note that during a path update,
the \emph{Packet-out} is be sent to the destination
reported in the \emph{Packet-in} which may
generate data plane traffic.
To prevent data plane traffic,
the malicious switch can use a
destination host that is connected to itself
(so that the \emph{Packet-out} is sent back to it). 
The message sequence pattern
for path update teleportation is shown in Figure~\ref{fig:hm_onos}.

We can summarize the scheme presented so far with the following abstract steps: A
switch~$s1$ announces~$X$, a switch~$s2$ announces~$X$ thereby stealing~$X$ 
from~$s1$, where stealing is detected by the ``victim" (~$s1$). 
Also note that announcing is possible once some host which is connected 
to the malicious switch (e.g.,~$k1$ at~$s1$) is learned by the controller. 

Based on these basic steps, we 
can generalize our scheme for~$m$ malicious switches. 
Each malicious switch,~$s_i$, with id~$i\in [m]$, should implement 
an event handler (see Algorithm~\ref{alg:pseudocode}),
in addition to the normal (non-malicious) behavior. We assume 
that all switches are programmed with the same list of~$m^2$ special MAC addresses~$\{X_{i,j}\}_{i,j\in[m]}$ (the \emph{pre-shared secrets}).
Note that once switch~$i$ discovers switch~$j$, 
it can contact it by sending packets with source~$X_{i,j}$ and destination address~$X_{j,i}$.

\begin{algorithm}[t]
\SetAlgoVlined
\DontPrintSemicolon
\SetKwSwitch{Reactor}{Event}{Comment}{process}{:}{on}{//}{end}

\small
\Reactor{incoming OpenFlow message}{
\BlankLine
\BlankLine
\Event{start teleportation}{
    - announce all~$\{X_{i,j}\}_{j\in[m]}$ and~$\{X_{j,i}\}_{j\in[m]}$\;
}
\Event{received \emph{Flow-delete} for~$X_{i,j}$ for some~$j\in[m]$}{
    - announce~$X_{i,j}$ \;
    - add~$j$ to~$Discovered\_Switches$ \;
}
}
\caption{Generalized pseudo-code executed by switch~$s_i$ to teleport information using path update.}
\label{alg:pseudocode}
\end{algorithm}

\textbf{Path Reset:}
We next discuss a second flow reconfiguration based
teleportation 
which we refer to as \emph{path reset}.
Recall that at the heart of any SDN controller lies
the functionality to set up host-to-host/network connectivity,
according to the network policy (e.g., defining constraints such
as bandwidth, link type and waypoints), which is translated
into device level configurations (e.g., flow rules). 
The ``pave-path technique'' is an 
attractive alternative to the hop-by-hop installation
of new flows: Once the controller receives a first 
\emph{Packet-in} event from some switch, it proactively updates the other
switches along the path.
 
In order to provide high availability, a controller also monitors the network state 
and makes necessary changes,
such as rerouting or resetting flows on switches,
when needed (e.g., due to a link failure).
For example, triggered by a \emph{Packet-in} event, a controller may learn
that (parts of) the path may no longer be available, and
hence initiates the reconfiguration/repair of the path.
We define the reinstallation of flows by the controller
on switches along a path as \emph{path reset}.
Accordingly, the path reset technique involves \emph{Packet-in}, \emph{Flow-mod} 
and \emph{Packet-out}
messages. 

\begin{figure}[t]
    \begin{center}
		\small
		\def\svgwidth{3.187in}
        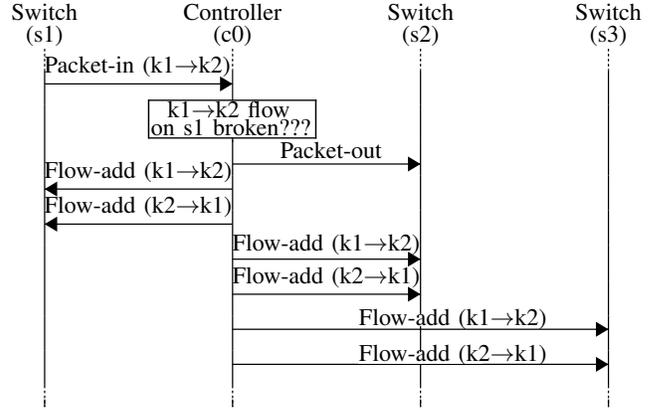
    \end{center}
    \caption{Message sequence pattern for path reset teleportation. Switch ~$s1$ teleports information
to~$s2$ and~$s3$ via \emph{Flow-add} messages sent by the controller~$c0$.}
    \label{fig:irf_onos}
\end{figure}

Malicious switches may use path reset
for implicit teleportation: If the controller resets the
complete path between hosts when it receives a \emph{Packet-in} from
a switch that ignores the flow rule, then
information can be communicated.
By doing this at multiple and specific times, 
a malicious switch can 
teleport information 
to other malicious switches 
along the path.
Figure~\ref{fig:irf_onos} illustrates
the message sequence pattern for teleportation using
path reset.

\subsection{Switch Identification}\label{sec:dpid}

This teleportation type exploits the fact that a switch 
typically must uniquely identify
itself whenever it connects to the controller. 
For example, in OpenFlow this is usually done 
using the Datapath-ID (DPID) field in the \emph{Features-reply} message.
We define two switches 
attempting to use the same DPID to connect to the same
logical controller as \emph{switch identification}.
The outcome can be used for
implicit teleportation.

Three basic ways an OpenFlow controller can react
to using the same DPID are as follows:
\begin{enumerate}
	\item The controller denies the second switch a connection.
	\item The controller terminates the first switch and connects to the second.
	\item The controller accepts both switches but sends them different \emph{Role-request} messages.
\end{enumerate}

\begin{figure}[t]
    \begin{center}
		\small
		\def\svgwidth{3.187in}
        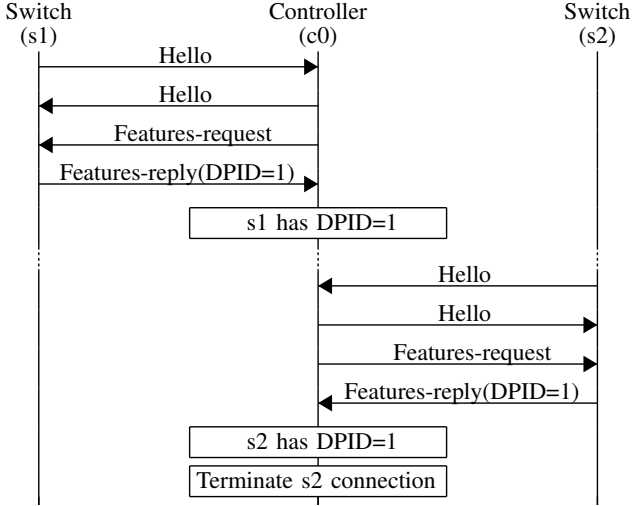
    \end{center}
    \caption{Message sequence pattern for \emph{switch identification} teleportation when the controller denies the second switch a connection. When~$s2$'s connection is terminated,~$s1$ successfully teleports information to~$s2$.}
    \label{fig:dpid_steal}
\end{figure}

With any of the above outcomes, a 
switch can infer the (mis)use of the same DPID by another switch. 
By using a-priori configured single or multiple 
DPID values, a pair of malicious switches can establish teleportation.
For example, consider the message sequence pattern in Figure~\ref{fig:dpid_steal}, 
and assume that first switch~$s1$ tells controller~$c0$
that its DPID is 1. At a later time, switch~$s2$ tells~$c0$ that its DPID is 1. 
At this point,
$c0$ does not allow~$s2$ to connect with DPID 1. 
Since~$c0$ denied~$s2$ to connect
with DPID 1,~$s1$ teleported 
information to
$s2$ via~$c0$.
With a similar message sequence pattern, 
the second outcome can be used for teleportation as well.  

Interestingly, switch identification is not limited to scenarios with a single
controller: We have found additional threats in the presence of \emph{distributed control planes}.
Moreover, we can generalize the first \emph{switch identification} outcome
to scenarios with~$m$ malicious switches,
see the event-handler algorithm, Algorithm~\ref{alg:pseudocode-dpid-onos}.
The other two outcomes discussed can also be
seen as event-handler algorithms.

\begin{figure}[t]
    \begin{center}
		\small
		\def\svgwidth{3.187in}
        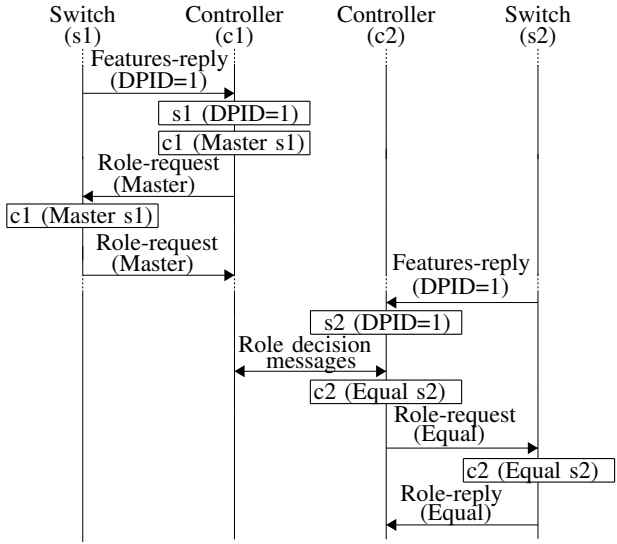
    \end{center}
    \caption{Message sequence pattern for \emph{switch identification} teleportation when controllers~$c1$ and~$c2$ send different Role-request messages to~$s1$ and~$s2$ respectively. When~$s1$ receives the \emph{Role-request=Master} message whereas~$s2$ receives the 
    \emph{Role-request=Equal} message. In this manner~$s1$ teleports information to~$s2$ when~$s2$ received the \emph{Role-request=Equal} message.}
    \label{fig:dpid_steal_mc}
\end{figure}




\begin{algorithm}[t]
\SetAlgoVlined
\DontPrintSemicolon
\SetKwSwitch{Reactor}{Event}{Comment}{process}{:}{on}{//}{end}

\small
\Reactor{connect to OpenFlow controller}{
\BlankLine
\Event{Features-request message from controller}{
    - announce DPID~$\{X_{i,j}\}_{j\in[m]}$ in \emph{Features-reply} message \;
}
\Event{Controller denies connection to announced DPID~$X_{i,j}$ for some~$j\in[m]$}{
    - add~$j$ to~$Discovered\_Switches$ \;
}
}
\caption{Generalized pseudo-code executed by switch~$s_i$ for \emph{switch identification} teleportation
when the controller denies the second switch a connection.}
\label{alg:pseudocode-dpid-onos}
\end{algorithm}

%
%
%


\subsection{Out-of-band Forwarding}\label{pktinpktout}

\begin{figure}[t]
    \begin{center}
		\small
		\def\svgwidth{3.187in}
        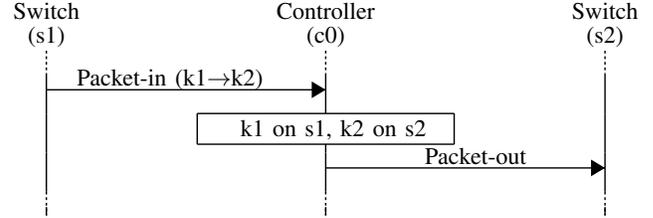
    \end{center}
    \caption{Message sequence pattern for \emph{out-of-band forwarding} teleportation. The controller~$c0$ receives the \emph{Packet-in} from~$s1$ and accordingly sends a \emph{Packet-out}
to~$s2$, successfully teleporting packets from~$s1$ to~$s2$.}
    \label{fig:oobfMsc}
\end{figure}

The third and potentially most
powerful teleportation technique is called \emph{out-of-band forwarding}.
It is an example of explicit teleportation.
Out-of-band forwarding exploits the fact that 
an SDN controller is typically connected to
multiple switches: Accordingly, a packet from one switch
can potentially reach multiple other switches in the network via the
control plane.
Out-of-band forwarding
involves a \emph{Packet-in} from one switch and a \emph{Packet-out} message 
at another switch, with
the possible side effect of \emph{Flow-mod} messages on the switch that sent
the \emph{Packet-in} message.
Out-of-band forwarding could for example include the
complete Ethernet frame (typically 1500 bytes),
and can even serve as a ``multicast service''.
Out-of-band forwarding can be a serious threat to network security,
not only because malicious traffic can bypass
critical security functions in the data plane,
but also because it can be exploited by switches and hosts.
Figure~\ref{fig:oobfMsc}, illustrates the message sequence pattern
for teleportation using out-of-band forwarding.

A summary of our teleportation techniques
is shown in Table~\ref{tab:building} along with
the type of teleportation 
and the associated OpenFlow messages.

\begin{table}[t]
\centering
\caption{Summary of teleportation techniques, types and associated threats.}
\label{tab:building}
\footnotesize
	\begin{tabular}{l l p{3.2cm}}
	\toprule
	\textbf{Technique}	& \textbf{Type}	 & \textbf{Threat}\\ \hline
	\midrule
	Flow (Re-)Configuration & Implicit & Covert communication and coordination.\\
	Switch Identification	& Implicit & Attack coordination.\\
	Out-of-band Forwarding & Explicit & Exfiltration, firewall/NIDS bypass and man-in-the-middle.\\
	\bottomrule
	\end{tabular}
\end{table}

\section{Switch- and Host-based Attacks}\label{sec:seceval}

We now demonstrate how the identified teleportation techniques 
can be exploited to carry out specific attacks.
In particular, we show how teleportation may be exploited:
\begin{enumerate}
\item To bypass security
critical network functions such as firewalls
and NIDSs;
\item As a rendezvous protocol for malicious switches;
\item To exfiltrate sensitive data from remote locations;
\item To conduct a man-in-the-middle (mitm) attack.
\end{enumerate}

Along the way, we also present a novel denial-of-service (dos) attack
(published as a CVE-2015-7516).

Before presenting the attacks in more detail,
we report on the setup we used to verify the attacks.

\subsection{Setup}

We verified all our attacks in 
a virtual machine, using Mininet-2.2.0 and
Open vSwitch-2.0.1 
for the data plane. For the control plane we used
ONOS-1.1.0 as it was the state-of-the-art.
At the time of our experimentation
Floodlight, OpenDaylight Lithium-SR2, and RYU v3.27,
still did not support the intent based framework.
Indeed our experiments showed that they were
only vulnerable to a subset of the attacks
(e.g., switch identification, out-of-band-forwarding\footnote{\url{https://goo.gl/FN9ULQ}})
For packet generation we use ping and nmap-6.40. 
We use \emph{ebtables} v2.0.10-4 (December 2011)
as our transparent firewall and \emph{Snort} version 2.9.6.0 GRE (Build 47) as our
NIDS. We modified code developed by 
\emph{austinmarton}~\cite{code-jumboframe} to
set the Ethertype field in an Ethernet frame. \emph{ettercap} 0.8.0 was used
with a custom HTTP filter for the mitm attack.

\subsection{Bypassing Critical Network Functions}

We believe that the possibility to bypass
network elements is a serious threat
in modern computer networks.
For example, many network policies today are defined in terms of adjacency 
matrices or big switch abstractions, specifying which traffic is allowed between an ingress
port~$s$ and an egress network port~$t$~\cite{bigswitch}. In order to enforce
such a policy, traffic from~$s$ to~$t$ needs to traverse a middlebox 
instance (waypoint) inspecting and classifying the 
flows. The location of every middlebox may be optimized, but is subject to
the constraint that the route from~$s$ to~$t$
should always go via the waypoint.

\subsubsection*{Firewall and NIDS}
\begin{figure}[t]
    \begin{center}
		\small
		\def\svgwidth{3.187in}
        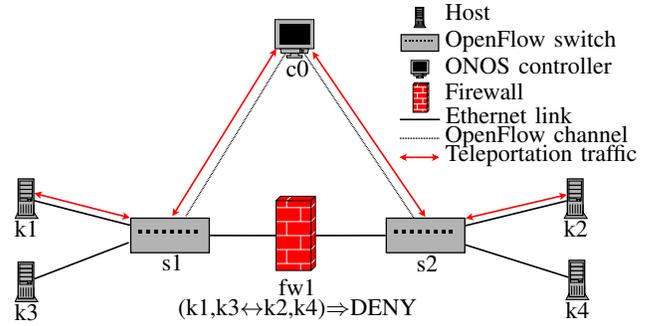
    \end{center}
    \caption{An SDN topology with OpenFlow switches~$s1$ and~$s2$ and an OpenFlow controller~$c0$ (ONOS).~$k1$ and~$k3$ are connected to~$s1$ while~$k2$ and~$k4$ are connected to~$s2$.~$s1$ and~$s2$ are separated by a firewall~$fw1$ that denies hosts on~$s1$ to communicate with hosts on~$s2$ and vice-versa.~$k1$ can use \emph{out-of-band forwarding teleportation} to transfer data to~$k2$, bypassing~$fw1$.}
    \label{fig:firewall_example}
\end{figure}

In order to demonstrate how a firewall may be circumvented
by hosts (or switches),
we set up Mininet and ONOS as shown in Figure~\ref{fig:firewall_example}.
The switches do not have flow rules for~$k1$ and~$k2$
to communicate.
The firewall~$fw1$ prevents hosts on the left to
communicate with hosts on the right and vice-versa.
ONOS has the \emph{Intent Reactive Forwarding} (\emph{ifwd}) application enabled.
\emph{ifwd} uses the reactive ``pave-path technique'' (discussed above) to install flows
in the switches.
By default, 
the \emph{ifwd} application establishes host-to-host connectivity when it receives
a \emph{Packet-in} for which no flows exist.

We send a ping packet from~$k1$ to~$k2$. Despite the presence of the
firewall,~$k1$ receives the reply from~$k2$ using out-of-band forwarding teleportation.
In the absence of out-of-band forwarding teleportation, the packet would have been
dropped by~$fw1$.

Indeed, in this case, out-of-band forwarding teleportation has the side effect of
installing flows on~$s1$ and~$s2$ for~$k1$ and~$k2$ to communicate,
preventing further out-of-band forwarding teleportation.
By masquerading its MAC address,~$k1$ can teleport more
data to~$k2$ via out-of-band forwarding teleportation.

Similar to the firewall scenario, 
we can also use out-of-band forwarding teleportation in the presence of 
\emph{Snort}, an NIDS.
In particular, we can generate 
attack traffic using nmap to conduct 
TCP flag attacks or even port scans. Indeed, by masquerading the source MAC 
address, one can effectively carry out a wide enough port scan
without having the scan pass through the firewall
and being detected by the latter.

By replacing the firewall we previously described with
\emph{Snort}, we use nmap from~$k1$ to carry out a TCP port scan on~$k2$
using out-of-band forwarding teleportation.
By inspecting the alerts in \emph{Snort} we verified that no alerts
were generated for the port scan.

Note that the host-to-host connectivity setup 
involves a \emph{Packet-in} and \emph{Flow-mod} messages whereas
 the out-of-band forwarding teleportation only involves \emph{Packet-in} and
 \emph{Packet-out} messages with the side effect of
\emph{Flow-mod} messages. 
Therefore, security policy 
enforcers that 
do not
inspect and correlate \emph{Packet-in} with \emph{Packet-out}s,
will miss out-of-band forwarding teleportation based attacks.
Of course, violating \emph{Flow-mod}s 
may eventually be detected, but only 
after the data has been teleported.





\subsection{Rendezvous and Malicious Switch Discovery}

We next consider a rendezvous protocol in which 
malicious switches
wish to discover one another. 
A rendezvous or discovery protocol
can be also seen as a precursor to a much
more damaging attack such as a denial-of-service,
man-in-the-middle (mitm) or exfiltration.

Teleportation can be  
an attractive solution: 
A rendezvous protocol can rely on steganography, i.e.,
embedding patterns in teleported benign information or
modulating patterns in legitimate messages.
Without teleportation,
by going through the data plane directly,
the malicious switches risk detection.

We show how three of our techniques, namely 
path update, 
path reset and switch identification
teleportation may be used
as a rendezvous protocol for
malicious switches.


\subsubsection{Path Update}\label{sec:disco_path_update}

\begin{figure}[t]
    \begin{center}
		\small
		\def\svgwidth{3.187in}
        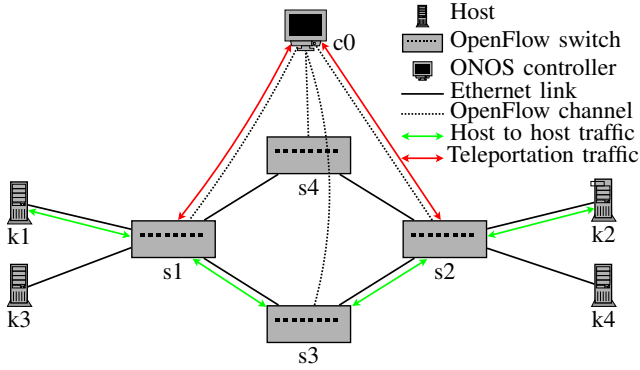
    \end{center}
    \caption{An SDN topology of OpenFlow switches~$s1$,~$s2$,~$s3$ and~$s4$, OpenFlow controller~$c0$ (ONOS). Hosts~$k1$ and~$k3$ are connected to~$s1$ and~$k2$ and~$k4$ are connected to~$s2$.~$c0$ has installed flows on~$s1$,~$s3$ and~$s2$ so that~$k1$ and~$k2$ can communicate bi-directionally. Teleportation traffic is via~$c0$.}
    \label{fig:onos}
\end{figure}

To demonstrate a rendezvous with path update teleportation, 
we set up Mininet and ONOS as shown in Figure~\ref{fig:onos}.
Instead of instrumenting code for the malicious switches, 
we keep them as simple Open vSwitches and we defined
dedicated Mininet hosts ($k3$ and~$k4$) for each of them. We use 
the dedicated hosts ($k3$ and~$k4$) to generate the packet that the malicious switch 
sends as a \emph{Packet-in} to the controller. 
The \emph{host mobility} and \emph{ifwd} applications are enabled on ONOS.
The controller has already installed flows for~$k1$ to~$k2$ and
vice-versa. 
Accordingly, we use~$k4$ connected to~$s2$, to 
send~$k2$ a packet using~$k1$ as the source MAC address.
This triggers the controller to issue \emph{Flow-mod} commands
to~$s1$,~$s2$ and~$s3$.~$s2$ thereby teleported its presence to~$s1$.

By inspecting the flows on the switches, we verified
the successful path update teleportation:~$s2$ was able
to cause a flow deletion in~$s1$ without exchanging 
any packets with~$s1$ directly (except for a normal flow in the past).

Note that path update may trigger alerts in systems
that keep track of moving MAC addresses by inspecting
\emph{Packet-in} and
\emph{Flow-mod} messages. In such cases, 
many
moving MAC addresses may introduce
suspicious activity within the network.
Also worth noting is that port-based
security (that associates MAC
addresses with specific ports)
may not be applicable in the presence
of malicious switches.

\subsubsection{Path Reset}\label{sec:disco_pathreset}

To demonstrate that path reset teleportation 
can be used as a rendezvous protocol, 
we consider the same setup as outlined in Section~\ref{sec:disco_path_update}.
We modulate traffic between~$k1$ and~$k2$ using ping 
packets with 100 microsecond intervals. 
Instead of manipulating the Open vSwitch code for sending a \emph{Packet-in} 
for an existing flow from~$s1$, we simply remove the flow 
for~$k1$ to~$k2$ on~$s1$, using the \texttt{ovs-ofctl del-flow} 
command. This causes~$s1$ to send~$c0$ a \emph{Flow-removed} message 
which triggers the controller to add the flow back onto~$s1$. 
But due to the high rate of ping traffic, at least one packet 
triggers a table-miss
before~$s1$ adds the flow and a \emph{Packet-in} 
is sent to~$c0$.

When~$c0$ receives the \emph{Packet-in}
it sends the packet to~$s2$ directly as a \emph{Packet-out}, bypassing~$s3$, 
and then sends \emph{Flow-mod}s to~$s1$,~$s2$ and~$s3$ resetting
the bi-directional path between~$k1$ and~$k2$. 
By checking the lifetime of the flow rules on~$s1$,~$s2$ and~$s3$
we verified that path reset teleportation succeeded.
In this manner,~$s1$ teleported its presence to~$s2$ by having
the controller send \emph{Flow-mod} commands for existing flow rules.

Note that such an attack works in the presence of
topology spoofing defenses~\cite{dhawan2015sphinx, sdn-visibility-poison} as the
\emph{Packet-in} and \emph{Flow-mod} messages generated do not
alter the existing topology.
Indeed, receiving a 
\emph{Packet-in} for a flow that exists in the switch is
suspicious but we are not aware of any work that
keeps track of such events. 



%
%

\subsubsection{Switch Identification}\label{sec:dpid}
We now demonstrate how two malicious switches may teleport their
presence using switch identification.
We set up Mininet and ONOS as shown in 
Figure~\ref{fig:onos} with only
$s1$,~$s3$ and~$s4$ having connected to~$c0$ with DPID 1, 3 and 4 respectively. 
Also, there are no flows installed on the switches for hosts to communicate.
We modified the Mininet script to configure~$s2$ with
the same DPID as~$s1$.

When~$s2$ tries to connect to~$c0$ with
DPID 1 after~$s1$ has connected to~$c0$, it is denied a connection. 
This way,
$s1$ teleports its presence to~$s2$.

In Floodlight and OpenDaylight,
when~$s2$ attempts to connect to~$c0$ with DPID 1 after~$s1$ has connected,
Floodlight terminates the connection with~$s1$
and accepts~$s2$'s connection.~$s2$ thereby teleports
its presence to~$s1$.
Interestingly RYU 
allowed switches with the same DPID to co-exist
which potentially introduces additional
issues.

\begin{figure}[t]
   \begin{center}
		\small
		\def\svgwidth{3.087in}
        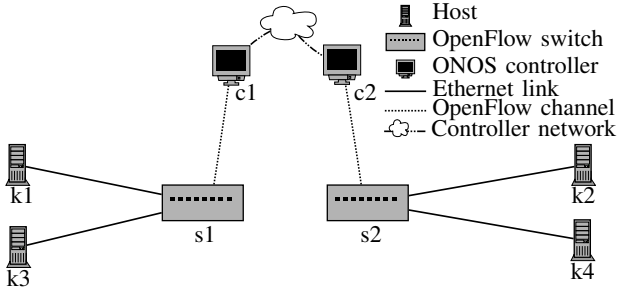
    \end{center}
    \caption{An SDN topology with independent OpenFlow switches controlled by independent OpenFlow controllers (ONOS).~$c1$ and~$c2$ share and synchronize state information via an independent controller network.~$s1$ is controlled by~$c1$ and~$s2$ is controlled by~$c2$ respectively.}
    \label{fig:dpid_steal_mc_topo}
\end{figure}

Switch identification teleportation is also possible when 
multiple controllers manage independent switches.
We set up Mininet and ONOS
as shown in Figure~\ref{fig:dpid_steal_mc_topo}.
Initially~$s1$ connects to~$c1$ with DPID 1.~$c1$ then declares itself as the \emph{Master}
for~$s1$. At a later time,~$s2$ connects to~$c2$ and claims to have DPID 1.
$c2$ then sends~$s2$ the \emph{Equal} role. In this manner,~$s1$
teleports its presence to~$s2$.
By inspecting the OpenFlow channels, we verified 
the different \emph{Role-request} messages sent by
the respective controllers to their respective switches.




\subsection{Exfiltration}\label{sec:exfil}
Our next attack is related to data exfiltration.
This is a key concern for many organizations that
own intellectual property, personal data or any kind of
sensitive information. Once an attacker gets into
a network, one possible goal of the attacker is to
stealthily
exfiltrate sensitive data.

We demonstrate exfiltration by 
considering a scenario where a small number of hosts
are networked together in a remote location.
The data plane isolation 
is meant to improve security.
However the data plane elements are managed by a controller that handles other
similar remote locations.
We show that in such a network,
not only malicious switches can exfiltrate data
using out-of-band forwarding teleportation but even
malicious hosts.

\begin{figure}[t]
    \begin{center}
		\small
		\def\svgwidth{3.187in}
        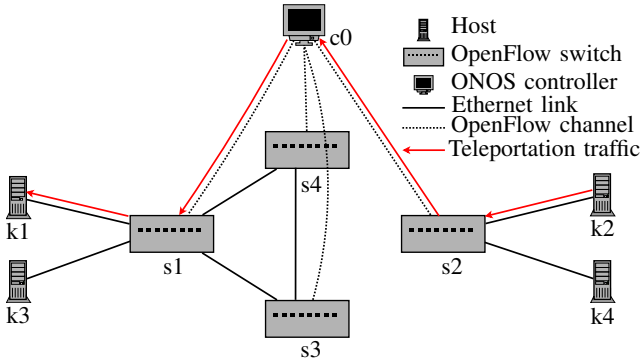
    \end{center}
    \caption{An SDN topology with OpenFlow switches~$s1$,~$s2$,~$s3$ and~$s4$ and an OpenFlow controller~$c0$ (ONOS).~$k1$ and~$k3$ are connected to~$s1$ while~$k2$ and~$k4$ are connected to~$s2$. Note that~$s2$ is not connected to the other switches, and thereby is isolated in the data plane.~$k2$ can still exfiltrate data to~$k1$ using \emph{out-of-band forwarding teleportation} circumventing the data plane isolation.}
    \label{fig:disc_switch}
\end{figure}

We set up Mininet and ONOS as shown in Figure~\ref{fig:disc_switch}.
ONOS has the \emph{ifwd} application activated.
By showing how~$k2$ can exfiltrate
data to~$k1$, we also demonstrate how~$s2$ can exfiltrate
data to~$k1$ or~$s1$.

Given that~$s1$ and~$s2$ do not have flow rules for
traffic from~$k2$ to~$k1$ (as they are located in
disconnected data planes),
$k2$ can exfiltrate data to~$k1$ by simply sending a packet
(e.g., UDP packet)
to~$k1$ thereby exploiting out-of-band forwarding teleportation.
The controller will receive the packet from~$s2$
and send it to~$s1$ which will then forward the
packet to~$k1$.

By inspecting the OpenFlow channels,
we can see the out-of-band forwarding teleportation, first
as a \emph{Packet-in} and then as a \emph{Packet-out}.


\subsection{Evading Policy Conflicts}

For an attacker, remaining stealthy is
key to persistent existence.
One of the side effects of using the
out-of-band forwarding teleportation is the \emph{Flow-mod} messages
issued by the controller.
The \emph{Flow-mod} messages may generate
policy conflicts (unauthorized/conflicting flow rules),
alerting the administrator.
A stealthier version of using the out-of-band forwarding teleportation
would be to prevent the \emph{Flow-mod} side effect.
This would not only prevent policy conflicts,
but also leave minimal traces on the source
and sink switches.

In order to demonstrate this attack, we set up
Mininet and ONOS with \emph{ifwd} activated as shown in
Figure~\ref{fig:disc_switch}. 
$k2$ can exfiltrate data to~$k1$ 
using out-of-band forwarding teleportation without triggering
\emph{Flow-mod}'s on~$s2$ and~$s1$ by
masquerading its source MAC address
\emph{and} ETHER\_TYPE (e.g., Jumbo frame: 0x8870).

\begin{figure}[t]
    \begin{center}
		\small
		\def\svgwidth{3.187in}
        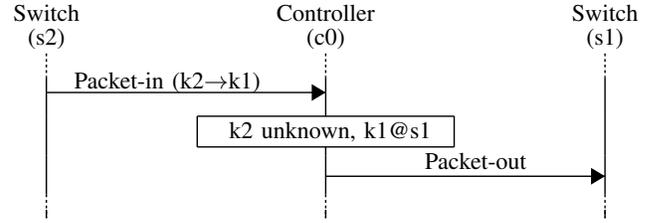
    \end{center}
    \caption{The message sequence pattern for evading policy conflicts using out-of-band forwarding teleportation. The side effect of \emph{Flow-mod} messages are avoided when Jumbo frames are used from a masqueraded MAC address; only \emph{Packet-in}s and \emph{Packet-out}s are used.}
    \label{fig:stealthy_irf}
\end{figure}

If the packet processor
and intent framework cannot correctly identify a packet,
their behavior may violate security policies.
Note that it is enough if the ETHER\_TYPE is set
to a value that ONOS does not recognize, and we are not
restricted to Jumbo frames only. The message sequence
pattern for out-of-band forwarding teleportation without the
\emph{Flow-mod} side effect is 
shown in Figure~\ref{fig:stealthy_irf}.

By inspecting the OpenFlow channels,
we can verify that the packet was indeed teleported
via out-of-band forwarding teleportation first as a \emph{Packet-in} and then
as a \emph{Packet-out}. By inspecting the flows
on the switches, we can verify that no new flows
are present.

\subsubsection*{Remark on a Denial-of-service Attack}

Interestingly, we observed that a side effect of our out-of-band forwarding teleportation  
is a novel denial-of-service attack.
If in our evading policy conflicts example,
the host sends the same packet (Jumbo frame) again, then \emph{ifwd}
encounters a null-pointer exception and 
disconnects the switch that sent it the packet.
This shows how a malicious host can cause the switch
it is connected to, to be disconnected from the
controller even when a packet it sends is not corrupted.

We emphasize
that this is a side effect of out-of-band forwarding teleportation only,
and not a teleportation issue in itself. 
Fortunately, the issue has been resolved by the ONOS
community after we contacted them 
(published as CVE-2015-7516).

\subsection{Man-In-The-Middle}

While we have so far focused on attacks where either only switches
or only hosts are malicious, we now detail an attack that involves a 
malicious switch and a malicious host.
The damage of such a collaboration
can be severe, for example, the attackers
could serve benign hosts with malicious web pages.
In order to exemplify the attack we use HTTP rather than HTTPS.

\begin{figure}[t]
    \begin{center}
		\small
		\def\svgwidth{3.187in}
        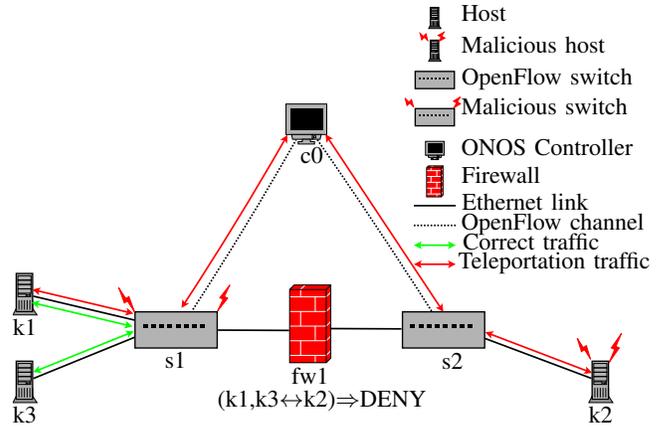
    \end{center}
    \caption{An SDN topology with OpenFlow switches~$s1$ and~$s2$ with~$c0$ the OpenFlow controller (ONOS).~$k1$ and~$k3$ are connected to~$s1$ while~$k2$ is connected to~$s2$.~$fw1$ denies~$k2$ to communicate with~$k1$ and~$k3$ and vice-versa via the data plane.~$s1$ and~$k2$ being malicious, exploit the out-of-band forwarding teleportation  to eavesdrop and modify communication data between~$k1$ and~$k3$ bypassing~$fw1$.}
    \label{fig:mitm_webserver}
\end{figure}

For this attack, we set up Mininet and ONOS with \emph{ifwd} activated
as shown in Figure~\ref{fig:mitm_webserver}.
$s1$ and~$k2$ are both malicious while the others are not.
$k3$ is a benign web server.~$s1$ teleports specific
HTTP traffic towards~$k2$.~$k2$ modifies the HTTP traffic
and teleports it back to~$s1$ who then forwards it to~$k1$.
In order to emulate the malicious switch, we introduced
a flow rule (shown in Listing~\ref{flowrule}) that rewrites the destination MAC address
for TCP traffic with PSH and ACK flags sent from~$k3$ to~$k1$, to~$k2$.
This modified packet is then passed through the flow table lookup again
by using the \texttt{resubmit} action in Open vSwitch.
$k2$ runs \emph{ettercap} to modify the TCP/HTTP payload
and forwards the packet to the correct destination. Specifically,
we created an \emph{ettercap} filter that looks inside HTTP responses
from~$k3$ for the word ``good", replaces it with ``evil", and sends
it to~$k1$.
The 
firewall~$fw1$ is meant to block traffic between hosts on the right
and the left.

When~$k1$ requests the \texttt{index.html} page from~$k3$,
based on the flow rule installed on~$s1$,
only HTTP responses from~$k3$ are teleported to~$s2$ and forwarded to~$k2$,
through the out-of-band forwarding teleportation. 
Subsequently,~$k2$ modifies only the \texttt{index.html}
web page and has~$s2$ teleport it back to~$s1$ via out-of-band forwarding teleportation.
Indeed, the side effect
is \emph{Flow-mod} messages to~$s1$ and~$s2$. 

By viewing the \texttt{index.html}
file received at~$k1$ we verified
that the mitm attack was successful.
The benign and malicious web pages
are shown in Listing~\ref{goodhtml} and Listing~\ref{badhtml}
respectively.
By inspecting the flow counters on the
switches we verified that necessary packets
did not pass through the data plane.

Note that we did not introduce code into the Open vSwitches
to handle the mitm, therefore
once the flows are installed on the switches,
the firewall will block all traffic between~$s1$ and~$s2$
and vice-versa.



\begin{lstlisting}[
    basicstyle=\footnotesize, %or \small or \footnotesize etc.
float=t,
frame=single, label=ettercap-filter, caption=An Open vSwitch flow rule that
was introduced into 
the malicious switch (\emph{$s1$}) to teleport HTTP traffic with the
PSH and ACK flags to the benign switch \emph{$s2$}. The matching
packets have the destination MAC address modified and resubmitted
to the flow-table lookup which results in \oobf{} teleportation.,
label=flowrule,
language=bash, linewidth=.95\columnwidth, captionpos=b,framexleftmargin=-3.4pt]
 priority=50001,tcp,in_port=2,
 dl_src=00:00:00:00:00:03,
 dl_dst=00:00:00:00:00:01,
 tp_src=80,tcp_flags=+psh+ack 
 actions=mod_dl_dst:00:00:00:00:00:02,
 resubmit:0
\end{lstlisting}

\begin{lstlisting}[
    basicstyle=\footnotesize, %or \small or \footnotesize etc.
float=t,frame=single, label=goodhtml, 
caption=HTML code from the benign web server. Note the word ``good'' is present in the body of the HTML code., 
label=goodhtml,
language=bash, linewidth=.95\columnwidth, rulecolor=\color{green}, captionpos=b,framexleftmargin=-3.4pt]
 root@Mininet-vm:~# curl http://10.0.0.2
 <html>
 <head>
 <title>Welcome page</title>
 <body>
 good
 </body>
 </html>
\end{lstlisting}

\begin{lstlisting}[
    basicstyle=\footnotesize, %or \small or \footnotesize etc.
float=t,frame=single, label=evilhtml, 
caption=HTML code modified by the malicious switch \emph{$s1$} and host \emph{$k2$}. Note the word ``evil'' is present in the body of the HTML code., 
label=badhtml,
language=bash, linewidth=.95\columnwidth, rulecolor=\color{red}, captionpos=b,framexleftmargin=-3.4pt]
 root@Mininet-vm:~# curl http://10.0.0.2
 <html>
 <head>
 <title>Welcome page</title>
 <body>
 evil
 </body>
 </html>
\end{lstlisting}



\section{Out-of-Band Forwarding Performance}
\label{sec:perf}

Having identified and demonstrated the various
attacks in this section we describe our evaluation
of the \emph{out-of-band forwarding} channel.
In particular we measure the throughput,
jitter and packet loss
of the channel, and
the resource footprint
of this channel
in terms of
CPU usage and
memory consumption at the controller.

\subsection{Setup}

In order to measure the throughput, jitter and packet loss of
the out-of-band forwarding channel, we
set up three dedicated
systems:
one system
(64 bit Intel Core i7-3517U CPU @ 1.90 GHz with
4GB of RAM)
running ONOS-1.5,
another system
(Intel Core 2 CPU @ 2.13 GHz with
4GB of RAM)
running
Mininet for the switches,
and a third system
(Intel Core i5-5200U CPU @ 2.20GHz with
16GB of RAM)
running OFCProbe~\cite{jarschel2014ofcprobe} for load generation.
Only the three systems are networked together via a Netgear 100Mbps
switch.
On the Mininet system, we use a simple line topology consisting of
two hosts and two switches, where,
host1 is connected to switch1 which is connected to switch2;
switch2 in turn is connected to host2.
The switches accordingly connect to ONOS as their controller.

\subsection{Methodology}
In order to emulate the malicious switch,
we simply install a flow rule on switch1 with
the highest priority so that the
out-of-band forwarding applies
by default, i.e., Packet-Ins are
sent to ONOS and forwarded accordingly.

We measure the throughput, jitter and packet loss using
iPerf3 running on the hosts in Mininet
using UDP packets.
We consider UDP packets with a payload of 512 bytes to be teleported.
Note that for the 512 bytes to be teleported, the overhead in bytes
for encapsulation (in the following order: Ethernet, IP, TCP, OpenFlow, Ethernet, IP, UDP)
is 110 bytes (for a Packet-in) and 108 bytes (for a Packet-out).
Therefore, a 10 Mbps teleportation channel
corresponds to approximately 2009 packets (Packet-ins) per second.
For the CPU and memory usage on the controller,
we use \emph{taskset} to pin ONOS to a single
CPU and use \emph{top} to measure the
CPU and memory usage.
For the load generation: OFCProbe,
emulates 20 switches that trigger
Packet-Ins to the controller following
a Poisson distribution ({$\lambda$=1}).
The throughput, jitter, packet-loss,
CPU and memory usage is sampled
every second for 600 seconds.



\subsection{Evaluation}

We first study the throughput of the 
UDP-based teleportation channel,
then consider the packet loss
and jitter characteristics,
and finally examine the resource footprint
in terms of CPU and memory in turn. 

\subsubsection{Throughput}

\begin{figure}[t]
    \begin{center}
        \includegraphics[width=0.99\columnwidth]{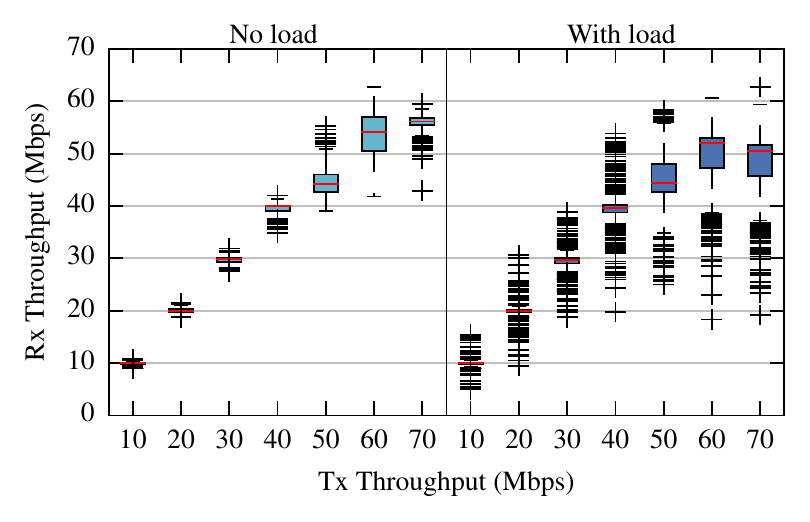}
    \end{center}
    \caption{Received throughput using \oobf{} without and with load on the controller.}
    \label{fig:throughput-10M-70M-10Trials}
\end{figure}

\begin{figure}[t]
    \begin{center}
        \includegraphics[width=0.99\columnwidth]{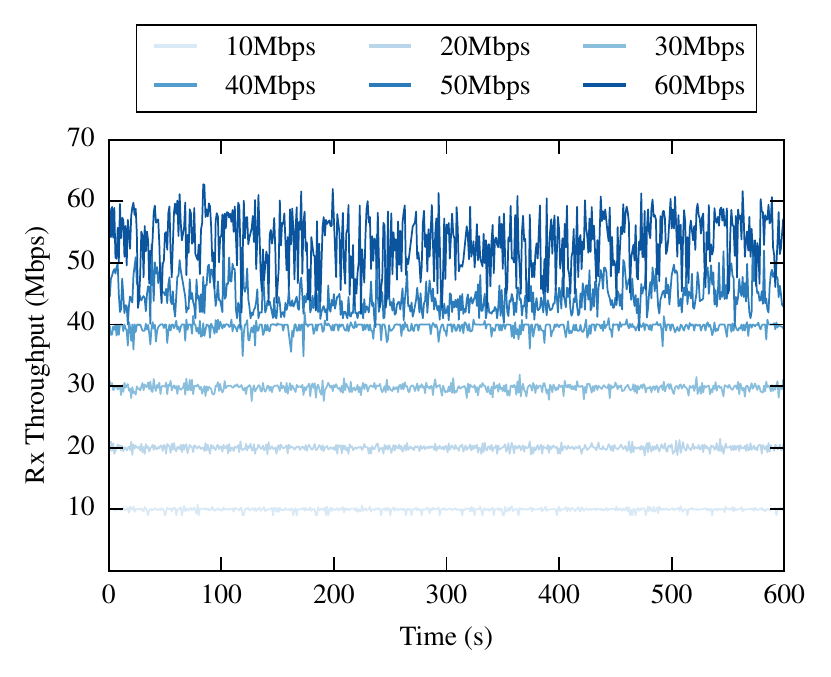}
    \end{center}
    \caption{Received throughput using \oobf{} without load on the controller.}
    \label{fig:throughput-10M-70M-10Trials-noLoad}
\end{figure}

\begin{figure}[t]
    \begin{center}
        \includegraphics[width=0.99\columnwidth]{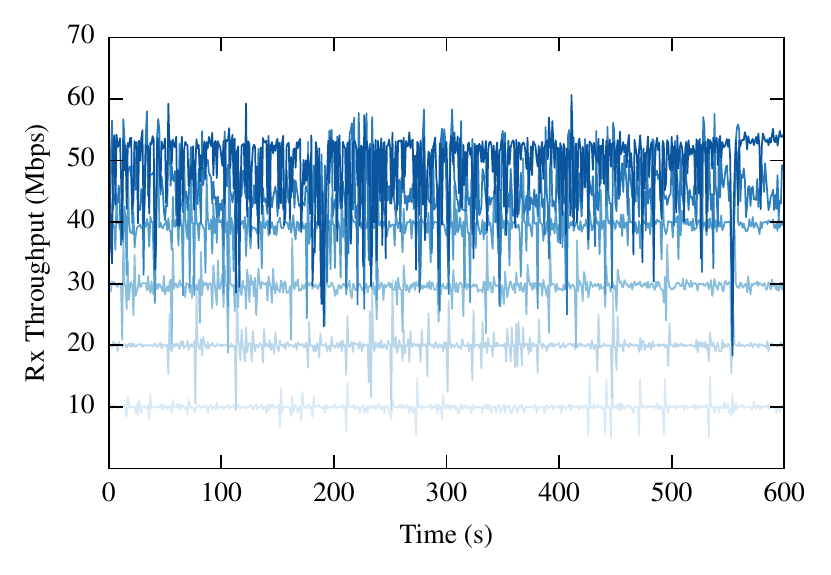}
    \end{center}
    \caption{Received throughput using \oobf{} with load on the controller. Legend as in Fig.~\ref{fig:throughput-10M-70M-10Trials-noLoad}}
    \label{fig:throughput-10M-70M-10Trials-withLoad}
\end{figure}

In Fig.~\ref{fig:throughput-10M-70M-10Trials} we
visualize the throughput of the teleportation channel
as box plots.
In Fig.~\ref{fig:throughput-10M-70M-10Trials-noLoad}~and~\ref{fig:throughput-10M-70M-10Trials-withLoad},
we visualize the throughput of the teleportation channel
without and with load resp.~as scatter plots.

We first observe that the teleportation channel can indeed sustain
very high transmission rates (Tx), of up to 40Mbps in both scenarios. 
In the scenario without load (Fig.~\ref{fig:throughput-10M-70M-10Trials}
and~\ref{fig:throughput-10M-70M-10Trials-noLoad}),
we see that the channel becomes saturated around rates slightly higher than 60Mbps,
after which the throughput suffers. 
In the scenario with load (Fig.~\ref{fig:throughput-10M-70M-10Trials}~and~\ref{fig:throughput-10M-70M-10Trials-withLoad}),
the variance of the throughput is naturally higher, but nevertheless it can sustain
rates which are almost as high as without load. 

In conclusion, our results show that the performance of teleportation
can go far beyond a small number of packets per second, 
which underlines the relevance (and potential threat)
of such channels.
 
\subsubsection{Packet Loss and Jitter}

\begin{figure}[t]
    \begin{center}
        \includegraphics[width=0.99\columnwidth]{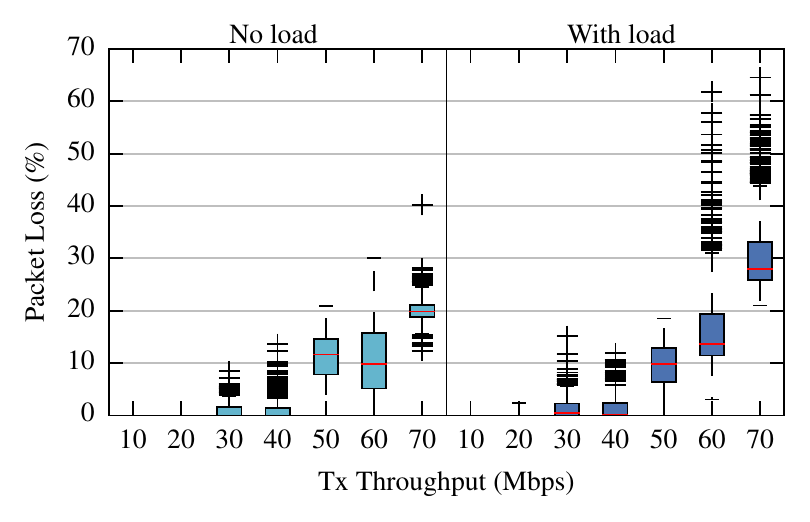}
    \end{center}
    \caption{Received packet loss using \oobf{} without and with no load on the controller.}
    \label{fig:packetLoss-10M-70M-10Trials}
\end{figure}

Fig.~\ref{fig:packetLoss-10M-70M-10Trials}
shows the packet loss for the scenario without
load and with load, respectively.
The experiments confirm the quality of the
considered teleportation channel: Up to 40 Mbps,
the packet loss is small despite some
variance, and naturally increases beyond 10\% for
rates more than 50 Mbps.
Again, only beyond the critical rates of slightly more than 50 Mbps
packet loss becomes significant.
Indeed, we can see a direct correlation between the packet loss
and the drop in throughput.

\begin{figure}[t]
    \begin{center}
        \includegraphics[width=0.99\columnwidth]{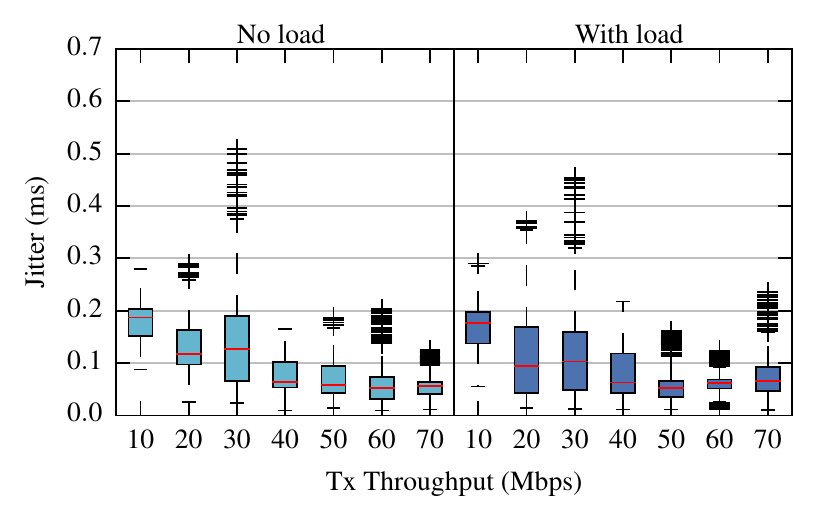}
    \end{center}
    \caption{Received jitter using \oobf{} without and with no load on the controller.}
    \label{fig:jitter-10M-70M-10Trials}
\end{figure}

We plot the jitter without load resp.~with load
in Fig.~\ref{fig:jitter-10M-70M-10Trials}
\footnote{Due to noise in our measurement setup,
we obtained some outliers in the jitter experiments.
Therefore, we followed the median absolute deviation~\cite{iglewicz1993detect}
method, with a tolerance of 3.5 to remove such outliers.}.
Also in terms of this metric, we can see that the teleportation
channel offers a good quality also for high rates.
The load on the controller again introduces some variance
to the jitter, however, it does not influence the median value by much.


\subsubsection{Resource Footprint}

To better understand the resource requirements
of the teleportation channel, as well as the reasons
behind the throughput drop at high rates,
we measured the CPU load and memory footprint on the controller.

\begin{figure}[t]
    \begin{center}
        \includegraphics[width=0.99\columnwidth]{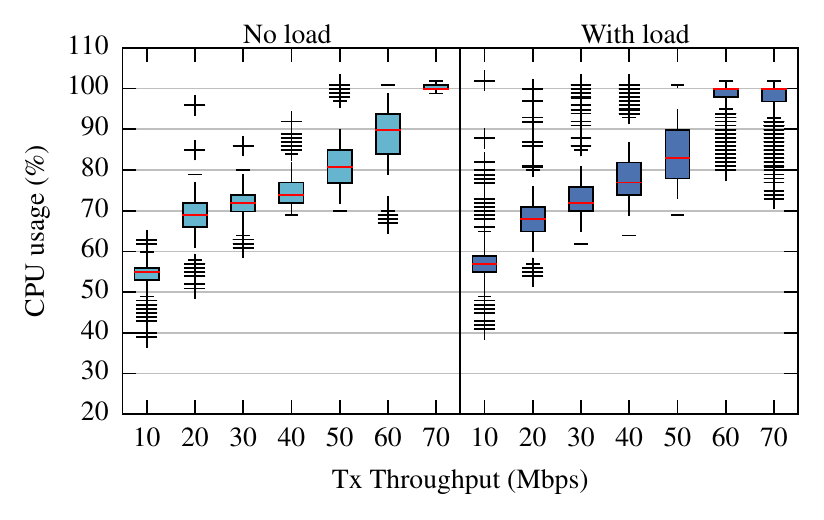}
    \end{center}
    \caption{CPU load using \oobf{} without and with no load on the controller.}
    \label{fig:cpu-10M-70M-10Trials}
\end{figure}

\begin{figure}[t]
    \begin{center}
        \includegraphics[width=0.99\columnwidth]{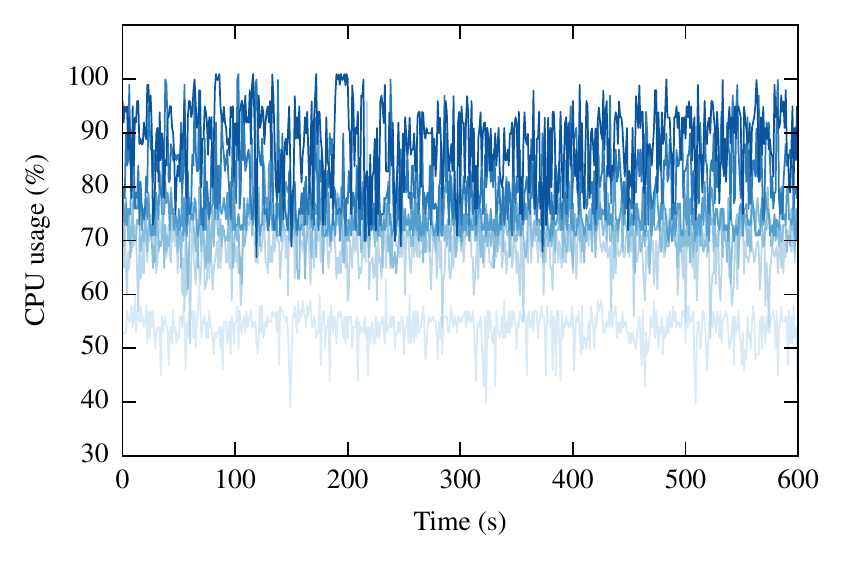}
    \end{center}
    \caption{CPU load using \oobf{} without load on the controller. Legend as in Fig.~\ref{fig:throughput-10M-70M-10Trials-noLoad}}
    \label{fig:cpu-10M-70M-10Trials-noLoad}
\end{figure}

\begin{figure}[t]
    \begin{center}
        \includegraphics[width=0.99\columnwidth]{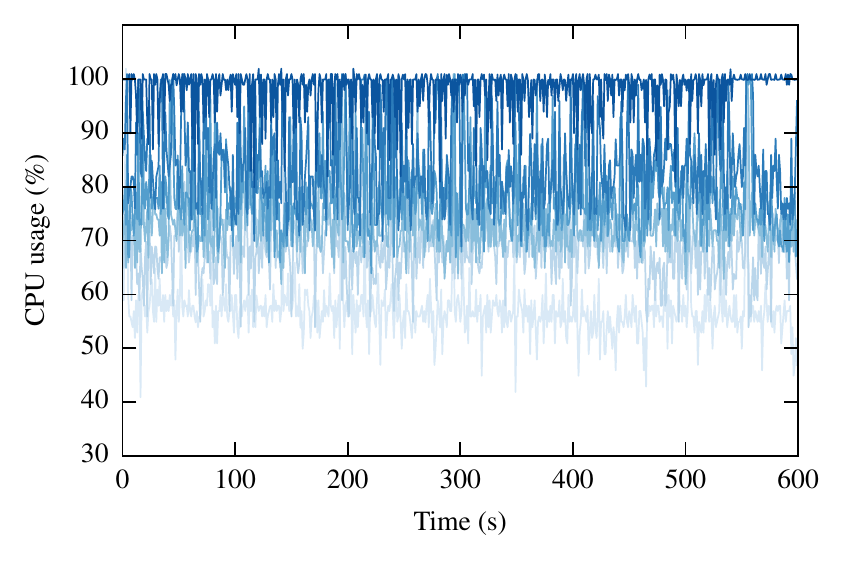}
    \end{center}
    \caption{CPU load using \oobf{} with load on the controller. Legend as in Fig.~\ref{fig:throughput-10M-70M-10Trials-noLoad}}
    \label{fig:cpu-10M-70M-10Trials-withLoad}
\end{figure}

Fig.~~\ref{fig:cpu-10M-70M-10Trials} visualizes
the CPU usage as a box plot, while
Fig.~\ref{fig:cpu-10M-70M-10Trials-noLoad}~and~\ref{fig:cpu-10M-70M-10Trials-withLoad}
visualizes the CPU loads over time. 
We observe that for a 10 Mbps channel, the CPU utilization
has a median value of 55, which is fairly high, but
not alarming. We also observe that at rates around 20 Mbps, the additional
CPU load introduced for an extra 10 Mbps (20 Mbps vs 30 Mbps channels) 
is small.
The influence of the load on the controller is discernible by
the variance introduced and a slight increase in the utilization.
However, again at around 50 Mbps, the effects become larger: 
We can clearly see the relationship
between the throughput and CPU load, and 
when the CPU consumption
begins to climb, the throughput begins
to drop.
Indeed, for transmission rates beyond 50 Mbps, the CPU utilization
is so high that it can easily be detected.
This is also the time around which
the jitter tends to increase by a small amount.

\begin{figure}[t]
    \begin{center}
        \includegraphics[width=0.99\columnwidth]{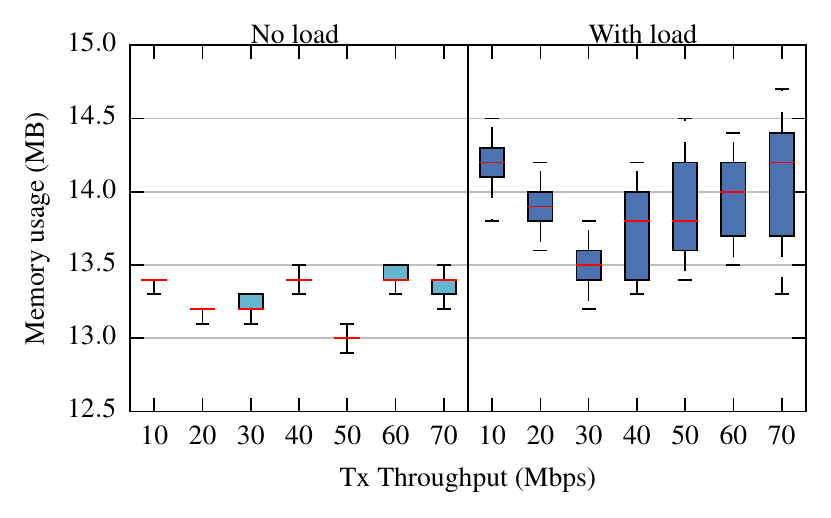}
    \end{center}
    \caption{Memory usage using \oobf{} without and with no load on the controller.}
    \label{fig:mem-10M-70M-10Trials}
\end{figure}

With respect to the memory consumption,
Fig.~\ref{fig:mem-10M-70M-10Trials}
shows that between 10 and 50 Mbps the memory consumption is
within a close range (13-15 MB) regardless of whether the 
load is induced or not. 
For 60Mbps and above, the memory consumption is higher.
Nonetheless, the impact of teleportation on the memory
is negligible.



\subsection{Summary}

Our first experiments show that teleportation channels
in the order of 10 Mbps are feasible, providing low packet loss
and low jitter. Moreover, these channels introduce
a moderate resource overhead in terms of CPU and a low overhead in
terms of memory.
Hence, such traffic may go unnoticed given the normal traffic patterns (even if
the regular traffic rate is orders of magnitude smaller).
However, we also observe that beyond a certain teleportation
rate, the CPU load will increase and become the bottleneck
for teleportation, limiting the throughput, introducing
high packet loss rates, and jitter. 
Therefore, we expect a sophisticated attacker to target
teleportation rates resulting in resource footprints which 
are obfuscated by
the regular load.

\section{Countermeasures}\label{sec:possible}

Having showcased a variety of attacks using teleportation, 
we now start exploring possible countermeasures.
Although we have demonstrated all the attacks using
ONOS 
we 
believe that these issues are 
likely to become
more general in nature.
They are becoming important with
the shift towards automated and
intent aware controller frameworks
allowing for simpler and agnostic
controller applications.
Based on our experiments we have also
seen that the resources required
and utilized for teleportation,
even at high rates are moderate. Therefore,
it may be difficult to distinguish
the attack traffic from the benign
traffic.
Accordingly we believe that,
with the separation of the control and data plane,
it is now important to monitor and police
the communication channel between the
separated planes due to the increased
attack surface~\cite{thimmaraju2016reigns}.
%




\subsection{Packet-in-Packet-out Watcher}
In order to prevent the out-of-band forwarding teleportation,
we strongly advise the use of a
\emph{Packet-in} and \emph{Packet-out} watcher.
It can either exist
as a controller application 
or as an application that
resides between the controller and
switches akin to hypervisors.
It would involve 
tracking and enforcing security policies for \emph{Packet-in}s
and their corresponding \emph{Packet-out}s.
Existing security enforcement kernels,
hypervisors and security applications
must account for \emph{Packet-in}s and \emph{Packet-out}s
in addition to \emph{Flow-mod}s to detect and
prevent out-of-band forwarding teleportation.

Note that the out-of-band forwarding teleportation
could also be used by malicious controller
applications. 
In a non-adversarial scenario, the order
in which a packet's fate is decided upon by
various applications can inadvertently
teleport the packet. Therefore,
verifying that the \emph{Packet-out} does
not reach an undesired switch/host can
prevent out-of-band forwarding teleportation.


\subsection{Audit-Trails and Accountability}
We propose controllers to introduce 
secure audit-trail capabilities, and accounting,
that enable network administrators to thoroughly
investigate events in their networks.
For example, controllers must log and alert
sensitive
events such as a moving MAC addresses, or,
receiving a \emph{Packet-in} when a flow has not
yet timed out. Such capabilities can aid
detection and prevention mechanisms.
It is also useful for
investigating 
security incidents.
We recommend administrators to frequently 
view controller logs, investigate failed events
and suspicious identities in the network.

\subsection{Enhanced IDS with Waypoint Enforcement}
Network intrusion detection systems
are an important 
means to detect and limit cyber attacks today,
and accordingly intrusion detection systems
constitute an
integral part of most networks.
We strongly suggest the use of an IDS application
on top of or before the controller, that can inspect
\emph{Packet-in}s and \emph{Packet-out}s and alert on suspicious traffic.
Indeed, some controllers today already offer 
basic functionality for waypoint enforcement.
In particular, we 
suggest waypoint enforcement and coordinating intrusion detection
systems from 
the control plane with the data plane.
This is non-trivial, but vital
for network security.

\section{Related Work}\label{sec:relwork}

While researchers have already pointed out several interesting
novel challenges in providing a correct operation
of networks with separate data and control planes~\cite{sharon,consistent-updates,correct-netvirt}, it is generally believed that SDN has the potential to render computer networking
more verifiable~\cite{header-space,veriflow} and even secure~\cite{anomal-traffic-sdn,netlord,fortnox,fresco,avantguard}.

Only recently researchers have started discovering
security threats in SDN. Kl{\"o}ti et al.~\cite{kloti-stride} 
report on a STRIDE threat analysis
of OpenFlow, and demonstrate  
data plane resource consumption attacks.
Kreutz et al.~\cite{Kreutz}
survey several threat vectors that may enable the exploitation of SDN vulnerabilities.
Benton et al.~\cite{Benton:2013:OVA:2491185.2491222}
analyze vulnerabilities in OpenFlow. In particular they point out the lack
of TLS adoption/implementation in OpenFlow switches and controllers.
In addition, they correctly identify the possibility of dos attacks
on the centralized control plane.
Another key challenge arising from the separation of the control and data planes,
is the potential loss of network visibility.
It has been shown that the network view of the controller
may even be poisoned~\cite{dhawan2015sphinx,sdn-visibility-poison}.
Thimmaraju et al.~\cite{thimmaraju2016reigns},
point out that threat models
for the virtualized
data plane need to account for a malicious/compromised data plane
in SDNs, and cloud operating systems such as OpenStack.

While much research went into designing more
robust and secure control planes~\cite{stn,nice},
less published work exists on the issue of malicious
switches. A notable exception is the work 
by Antikainen et al.~\cite{spook}, 
who consider the possibility of a malicious relay node 
for a man-in-the-middle attack. Interestingly, in our paper,
 we have shown that the
relay node can be the benign controller itself.
%
%
%

To the best of our knowledge,
our work is the first to point out and characterize the fundamental 
problem of SDN teleportation.
More generally, while most prior studies about malicious
switches focus on (indirect) \emph{attacks targeting the controller},
we in this paper demonstrate new kinds of attacks which
merely exploit the controller for \emph{directly} attacking (e.g., the confidentiality or availability)
of network services.

%

However, there are a number of interesting approaches
proposed in the literature which have implications for
our scenarios as well.
For example, the pre- and post-conditions of
Topoguard~\cite{sdn-visibility-poison} 
can defend
against our path update attack. However,
if the switches are malicious, these conditions
can be spoofed by the malicious switches. Also,
Topoguard 
cannot detect teleportation using path reset,
switch identification and out-of-band forwarding teleportation.

Sphinx~\cite{dhawan2015sphinx} can alert on the
path update teleportation. However,
it cannot detect
the path reset as the flow graph 
remains the same. Additionally,
Sphinx assumes that
switches cannot use the same DPIDs, therefore,
we believe that our switch identification teleportation
will not be detected by Sphinx.
Also, our out-of-band forwarding relies on \emph{Packet-in} and \emph{Packet-out} messages,
while \emph{Packet-out}s are not considered by Sphinx\footnote{Unfortunately,
the source code of Sphinx is not available.}. Therefore the 
suggested out-of-band forwarding teleportation can evade Sphinx,
until topology altering flows are installed.

Porras et al.~\cite{porras2015securing} propose
a security mediator that comprises of 
Rule Conflict Analysis,
Role-based Source Authentication,
State Table Manager and
a Permission Mediator.
We admit that the path update
can be detected using this approach, however,
our path reset does not introduce any
conflicting rules.
The \emph{Features-reply} messages are
not a part of their solution, therefore,
we believe that \emph{switch identification} teleportation
can succeed.
With respect to out-of-band forwarding teleportation,
unless the mediator investigates
the destination switch or MAC address
in the \emph{Packet-out},
the teleportation
can bypass the
security mediator given sufficient
permissions. 

SDN Hypervisors such as CoVisor~\cite{jin2015covisor}, 
Flowvisor~\cite{sherwood2009flowvisor}, FortNOX~\cite{fortnox}
depend on policies maintained in the hypervisor.
Therefore, we believe that all our teleportation
mechanisms hold unless a specific policy blocks
it.

Dover Networks~\cite{doverdenial} discovered the
behavior of Floodlight with switches using the same
DPID,  which we exploit for teleportation.

While \emph{Security-Mode ONOS}~\cite{sonos}
can enhance the security in many scenarios, by
introducing roles and permissions,
at least today, it does not help against teleportation:
Once \emph{ifwd} has the permission
to write intents and emit packets,
our teleportation succeeds. These
permissions are bare necessities for \emph{ifwd}
to function.


\begin{table}[t]
\centering
\caption{Summary of teleportation attacks and involved entities.}
\label{tab:teleattacks}
\footnotesize
	\begin{tabular}{p{2.4cm} p{2.8cm} p{2.0cm}}
	\toprule
	\textbf{Attack}	& \textbf{Teleportation technique}	 & \textbf{Exploited by}\\ \hline
	\midrule
	Bypass Firewall	&Out-of-band forwarding&Switch and Host\\
	Bypass NIDS &Out-of-band forwarding&Switch and Host\\
	Exfiltration&Out-of-band forwarding&Switch and Host\\
	Evading policy conflicts&Out-of-band forwarding&Switch and Host\\
	Man-in-the-middle&Out-of-band forwarding&Switch and Host\\
	\multirow{3}{*}{Rendezvous}&
	Path update&Switch\\
	&Path reset&Switch\\
	&Switch identification&Switch\\
	\bottomrule
	\end{tabular}
\end{table}

\section{Conclusions}\label{sec:conclusion}

As OpenFlow networks transition 
from research to production, new levels of 
reliability and performance are necessary~\cite{need-to-know}. 
This paper has identified and demonstrated  
a novel security threat introduced by software-defined
networks separating the control plane from the data plane.
In the presence of an unreliable south-bound interface
(containing malicious switches):
We have shown that state-of-the-art controller(s)
are vulnerable to teleportation. Teleportation
has numerous applications (cf.~the summary in Table~\ref{tab:teleattacks}): It 
can be exploited 
to bypass security-critical network elements (e.g.,
to exfiltrate confidential information),
as a discovery protocol for malicious switches,
to evade policy conflicts as well as for man-in-the-middle
attacks. 
Based on our preliminary evaluation, we can 
say that even a teleportation channel of over 10 Mbps
can easily be used inside a loaded control channel.


Our work can also be seen as a first security analysis
of the increasingly popular 
intent-based network mechanisms: 
While intent-based mechanisms are attractive
for allowing (cloud) network
operators resp.~SDN applications to focus on 
``what to connect'' rather
than ``how', we have shown that controller managed
intents need to be used with care. 
Indeed, our experiments with controllers
that are only starting to introduce an
intent based mechanism are not yet
vulnerable to all the specific attacks
presented in this paper.
Moreover, while intent mechanism implementations
can vary across controllers, we believe that the
underlying issues are fundamental. 

We understand our work as a first step, and believe
that our paper opens several relevant directions
for future research. In particular, we plan to extend our 
vulnerability analysis
to other SDN protocols and
conduct a more in-depth performance analysis. 
Another relevant avenue for future research regards
the development of countermeasures. 


\section*{Acknowledgments}
Research (partially) supported by the Danish Villum project
\emph{ReNet}.
We would like to thank the anonymous reviewers of Euro S\&P'17 for their valuable
feedback and helpful comments during the review process.
We would also like to thank Prof.~Jean-Pierre Seifert and
Tobias Fiebig for their constructive inputs at the onset of
this paper.

\bibliographystyle{IEEEtran}
\balance
\bibliography{literature}

\begin{thebibliography}{10}
\providecommand{\url}[1]{#1}
\csname url@samestyle\endcsname
\providecommand{\newblock}{\relax}
\providecommand{\bibinfo}[2]{#2}
\providecommand{\BIBentrySTDinterwordspacing}{\spaceskip=0pt\relax}
\providecommand{\BIBentryALTinterwordstretchfactor}{4}
\providecommand{\BIBentryALTinterwordspacing}{\spaceskip=\fontdimen2\font plus
\BIBentryALTinterwordstretchfactor\fontdimen3\font minus
  \fontdimen4\font\relax}
\providecommand{\BIBforeignlanguage}[2]{{%
\expandafter\ifx\csname l@#1\endcsname\relax
\typeout{** WARNING: IEEEtran.bst: No hyphenation pattern has been}%
\typeout{** loaded for the language `#1'. Using the pattern for}%
\typeout{** the default language instead.}%
\else
\language=\csname l@#1\endcsname
\fi
#2}}
\providecommand{\BIBdecl}{\relax}
\BIBdecl

\bibitem{eu}
\BIBentryALTinterwordspacing
{Internet Science Working Group}, ``Internet as a critical infrastructure:
  Security, resilience and dependability aspects ({JRA7}),'' 2015, accessed:
  2017-02-06. [Online]. Available:
  \url{http://www.internet-science.eu/groups/internet-critical-infrastructure-security-resilience-and-dependability-aspects}
\BIBentrySTDinterwordspacing

\bibitem{cheap}
{M. Armbrust et al.}, ``A view of cloud computing,'' \emph{Communications of
  the ACM}, pp. 50--58, April 2010.

\bibitem{ossi}
T.~Anderson, L.~Peterson, S.~Shenker, and J.~Turner, ``Overcoming the internet
  impasse through virtualization,'' \emph{IEEE Computer}, vol.~38, no.~4, pp.
  34--41, April 2005.

\bibitem{psbgp}
P.~v. Oorschot, T.~Wan, and E.~Kranakis, ``On interdomain routing security and
  pretty secure bgp (psbgp),'' \emph{ACM Trans. on Information and System
  Security (TISSEC)}, no.~3, July 2007.

\bibitem{header-space}
P.~Kazemian, G.~Varghese, and N.~McKeown, ``Header space analysis: Static
  checking for networks,'' in \emph{Proc. NSDI}, 2012, pp. 113--126.

\bibitem{veriflow}
A.~Khurshid, X.~Zou, W.~Zhou, M.~Caesar, and P.~B. Godfrey, ``Veriflow:
  Verifying network-wide invariants in real time,'' in \emph{Proc. NSDI}, 2013,
  pp. 467--472.

\bibitem{sharon}
O.~Padon, N.~Immerman, A.~Karbyshev, O.~Lahav, M.~Sagiv, and S.~Shoham,
  ``Decentralizing sdn policies,'' in \emph{Proc. ACM Principles of Programming
  Languages (POPL)}, 2015.

\bibitem{consistent-updates}
M.~Reitblatt, N.~Foster, J.~Rexford, and D.~Walker, ``Consistent updates for
  software-defined networks: Change you can believe in!'' in \emph{Proc. ACM
  Workshop on Hot Topics in Networks (HotNETs)}, 2011, pp. 7:1--7:6.

\bibitem{netlord}
J.~Mudigonda, P.~Yalagandula, J.~Mogul, B.~Stiekes, and Y.~Pouffary, ``Netlord:
  A scalable multi-tenant network architecture for virtualized datacenters,''
  in \emph{Proc. ACM SIGCOMM}, 2011, pp. 62--73.

\bibitem{fortnox}
P.~Porras, S.~Shin, V.~Yegneswaran, M.~Fong, M.~Tyson, and G.~Gu, ``A security
  enforcement kernel for {OpenFlow} networks,'' in \emph{Proc. ACM Workshop on
  Hot Topics in Software Defined Networking (HotSDN)}, 2012, pp. 121--126.

\bibitem{fresco}
S.~Shin, P.~Porras, V.~Yegneswaran, M.~Fong, G.~Gu, and M.~Tyson, ``Fresco:
  Modular composable security services for software-defined networks,'' in
  \emph{Proc. NDSS}, 2013.

\bibitem{avantguard}
S.~Shin, V.~Yegneswaran, P.~Porras, and G.~Gu, ``{AVANT-GUARD}: Scalable and
  vigilant switch flow management in software-defined networks,'' in
  \emph{Proc. ACM Conference on Computer and Communications Security (CCS)},
  2013, pp. 413--424.

\bibitem{thimmaraju2016reigns}
K.~Thimmaraju \emph{et~al.}, ``Reins to the cloud: Compromising cloud systems
  via the data plane,'' \emph{arXiv preprint arXiv:1610.08717}, 2016.

\bibitem{snowdencisco}
\BIBentryALTinterwordspacing
``Snowden: The {NSA} planted backdoors in {Cisco} products,'' 2014, accessed:
  02-01-2018. [Online]. Available:
  \url{http://www.infoworld.com/article/2608141/internet-privacy/snowden--the-nsa-planted\\-backdoors-in-cisco-products.html}
\BIBentrySTDinterwordspacing

\bibitem{1001dsl}
\BIBentryALTinterwordspacing
``The tale of one thousand and one dsl modems,'' 2012, accessed: 2017-02-06.
  [Online]. Available:
  \url{https://securelist.com/analysis/publications/57776/the-tale-of-one-thousand-and-one-dsl-modems/}
\BIBentrySTDinterwordspacing

\bibitem{synful}
\BIBentryALTinterwordspacing
``Synful knock - a cisco router implant - part i,'' 2015, accessed: 2017-02-06.
  [Online]. Available:
  \url{https://www.fireeye.com/blog/threat-research/2015/09/synful_knock_-_acis.html}
\BIBentrySTDinterwordspacing

\bibitem{lindner2009cisco}
\BIBentryALTinterwordspacing
F.~Lindner, ``Cisco {IOS} router exploitation,'' Recurity Labs, Tech. Rep.,
  2009, accessed: 2017-02-06. [Online]. Available:
  \url{http://www.blackhat.com/presentations/bh-usa-09/LINDNER/BHUSA09-Lindner-RouterExploit-PAPER.pdf}
\BIBentrySTDinterwordspacing

\bibitem{lee2006secure}
S.~Lee, T.~Wong, and H.~S. Kim, ``Secure split assignment trajectory sampling:
  A malicious router detection system,'' in \emph{Proc. IEEE/IFIP Transactions
  on Dependable and Secure Computing (DSN)}, 2006, pp. 333--342.

\bibitem{huawei}
\BIBentryALTinterwordspacing
``Huawei hg8245 backdoor and remote access,'' 2013, accessed: 2017-02-06.
  [Online]. Available:
  \url{http://websec.ca/advisories/view/Huawei-web-backdoor-and-remote-access}
\BIBentrySTDinterwordspacing

\bibitem{netisbackdoor}
\BIBentryALTinterwordspacing
``Netis routers leave wide open backdoor,'' 2014, accessed: 2017-02-06.
  [Online]. Available:
  \url{http://blog.trendmicro.com/trendlabs-security-intelligence/netis-routers-leave-wide-open-backdoor/}
\BIBentrySTDinterwordspacing

\bibitem{inception}
\BIBentryALTinterwordspacing
S.~Fagerland, W.~Grange, and {Blue Coat Systems Inc.}, ``The inception
  framework: Cloud-hosted {APT},'' 2014, accessed: 2017-02-06. [Online].
  Available: \url{http://dc.bluecoat.com/Inception_Framework}
\BIBentrySTDinterwordspacing

\bibitem{onos}
``{ONOS} wiki home,''
  \url{https://wiki.onosproject.org/display/ONOS/Wiki+Home}, 2017, accessed:
  02-01-2018.

\bibitem{onosDos}
\BIBentryALTinterwordspacing
{ONOS}, ``Security advisories,'' 2015, accessed: 2017-02-06. [Online].
  Available:
  \url{https://wiki.onosproject.org/display/ONOS/Security+advisories}
\BIBentrySTDinterwordspacing

\bibitem{specification2013version}
\emph{{OpenFlow Switch Specification}}, {Open Networking Foundation}, 2013,
  version 1.3.2 Wire Protocol 0x04.

\bibitem{code-jumboframe}
\BIBentryALTinterwordspacing
``Send a raw ethernet frame in linux.'' 2011, accessed: 2017-02-06. [Online].
  Available: \url{https://gist.github.com/austinmarton/1922600}
\BIBentrySTDinterwordspacing

\bibitem{bigswitch}
N.~Kang, Z.~Liu, J.~Rexford, and D.~Walker, ``Optimizing the "one big switch"
  abstraction in software-defined networks,'' in \emph{Proc. ACM CoNEXT}, 2013.

\bibitem{dhawan2015sphinx}
M.~Dhawan, R.~Poddar, K.~Mahajan, and V.~Mann, ``Sphinx: Detecting security
  attacks in software-defined networks.'' in \emph{Proc. NDSS}, 2015.

\bibitem{sdn-visibility-poison}
S.~Hong, L.~Xu, H.~Wang, and G.~Gu, ``Poisoning network visibility in
  software-defined networks: New attacks and countermeasures,'' in \emph{Proc.
  NDSS}, 2015.

\bibitem{jarschel2014ofcprobe}
M.~Jarschel \emph{et~al.}, ``Ofcprobe: A platform-independent tool for openflow
  controller analysis,'' in \emph{Proc. IEEE International Conference on
  Communications and Electronics}.\hskip 1em plus 0.5em minus 0.4em\relax IEEE,
  2014, pp. 182--187.

\bibitem{iglewicz1993detect}
B.~Iglewicz and D.~C. Hoaglin, \emph{How to detect and handle outliers}.\hskip
  1em plus 0.5em minus 0.4em\relax Asq Press, 1993, vol.~16.

\bibitem{correct-netvirt}
W.~Zhou, D.~K. Jin, J.~Croft, M.~Caesar, and P.~B. Godfrey, ``Enforcing
  customizable consistency properties in software-defined networks,'' in
  \emph{Proc. NSDI}, 2015, pp. 73--85.

\bibitem{anomal-traffic-sdn}
S.~A. Mehdi, J.~Khalid, and S.~A. Khayam, ``Revisiting traffic anomaly
  detection using software defined networking,'' in \emph{Proc. RAID Recent
  Advances in Intrusion Detection}, 2011, pp. 161--180.

\bibitem{kloti-stride}
R.~Kl{\"o}ti, V.~Kotronis, and P.~Smith, ``Openflow: A security analysis,'' in
  \emph{Proc. IEEE International Conference on Network Protocols (ICNP)}, Oct
  2013, pp. 1--6.

\bibitem{Kreutz}
D.~Kreutz, F.~M. Ramos, and P.~Verissimo, ``Towards secure and dependable
  software-defined networks,'' in \emph{Proc. ACM Workshop on Hot Topics in
  Software Defined Networking (HotSDN)}, 2013, pp. 55--60.

\bibitem{Benton:2013:OVA:2491185.2491222}
K.~Benton, L.~J. Camp, and C.~Small, ``Openflow vulnerability assessment,'' in
  \emph{Proc. ACM Workshop on Hot Topics in Software Defined Networking
  (HotSDN)}, 2013, pp. 151--152.

\bibitem{stn}
M.~Canini, P.~Kuznetsov, D.~Levin, and S.~Schmid, ``A distributed and robust
  sdn control plane for transactional network updates,'' in \emph{Proc. IEEE
  INFOCOM}, 2015, pp. 190--198.

\bibitem{nice}
M.~Canini, D.~Venzano, P.~Pere\v{s}\'{\i}ni, D.~Kosti\'{c}, and J.~Rexford, ``A
  nice way to test openflow applications,'' in \emph{Proc. NSDI}, vol.~12,
  2012, pp. 127--140.

\bibitem{spook}
M.~Antikainen, T.~Aura, and M.~S{\"a}rel{\"a}, ``Spook in your network:
  Attacking an sdn with a compromised openflow switch,'' in \emph{Proc. Secure
  IT Systems: Nordic Conf.}\hskip 1em plus 0.5em minus 0.4em\relax Springer
  International Publishing, 2014, pp. 229--244.

\bibitem{porras2015securing}
P.~Porras, S.~Cheung, M.~Fong, K.~Skinner, and V.~Yegneswaran, ``Securing the
  software-defined network control layer,'' in \emph{Proc. NDSS}, 2015.

\bibitem{jin2015covisor}
X.~Jin, J.~Gossels, J.~Rexford, and D.~Walker, ``Covisor: A compositional
  hypervisor for software-defined networks,'' in \emph{Proc. NSDI}, 2015.

\bibitem{sherwood2009flowvisor}
R.~Sherwood, G.~Gibb, K.-K. Yap, G.~Appenzeller, M.~Casado, N.~McKeown, and
  G.~Parulkar, ``Flowvisor: A network virtualization layer,'' OpenFlow, Tech.
  Rep., 2009.

\bibitem{doverdenial}
\BIBentryALTinterwordspacing
J.~M. Dover, ``A denial of service attack against the open floodlight sdn
  controller,'' Dover Networks, Tech. Rep., 2013. [Online]. Available:
  \url{http://dovernetworks.com/wp-content/uploads/2013/12/OpenFloodlight-12302013.pdf}
\BIBentrySTDinterwordspacing

\bibitem{sonos}
\BIBentryALTinterwordspacing
``{Security-Mode ONOS},'' 2015, accessed: 2017-02-06. [Online]. Available:
  \url{https://wiki.onosproject.org/display/ONOS/Security-Mode+ONOS}
\BIBentrySTDinterwordspacing

\bibitem{need-to-know}
M.~Kuźniar, P.~Perešíni, and D.~Kostić, ``What you need to know about sdn
  flow tables,'' in \emph{Proc. Passive and Active Measurement (PAM)}, 2015,
  pp. 347--359.

\end{thebibliography}

\end{document}